\documentclass[journal]{IEEEtran}
\usepackage[utf8]{inputenc}
\usepackage{amsmath}
\usepackage{amsthm}
\usepackage{amssymb}
\usepackage{graphicx}
\usepackage{cite}
\usepackage{epstopdf}
\usepackage{float}

\usepackage{algorithmicx}
\usepackage{amsfonts}
\usepackage{subcaption}
\usepackage[noend]{algpseudocode}
\usepackage[ruled,norelsize,linesnumbered]{algorithm2e}
\usepackage{multicol}

\newtheorem{theorem}{Theorem}[section]

\newtheorem{remark}[theorem]{Remark}

\makeatletter
\newcommand{\removelatexerror}{\let\@latex@error\@gobble}
\g@addto@macro{\@algocf@init}{\SetKwInOut{Parameter}{Parameters}}
\makeatother

\algdef{SE}[DOWHILE]{Do}{doWhile}{\algorithmicdo}[1]{\algorithmicwhile\ #1}%
\makeatletter
\def\BState{\State\hskip-\ALG@thistlm}

\newlength\myindent
\title{\Large\bf A New Microscopic Traffic Model Using a Spring-Mass-Damper-Clutch System}
\author{Zhaojian~Li,
  Firas~Khasawneh,
  Xiang~Yin,
  Aoxue~Li,
  Ziyou~Song
\thanks{\quad Zhaojian Li, Firas Khasawneh, and Aoxue Li are with the Department of Mechanical Engineering, Michigan State University, East Lansing, MI 48824, USA.
        Email: {\tt\small \{lizhaoj1,khasawn3,liaoxue\}@egr.msu.edu}}
\thanks{\quad Xiang Yin is with the Department of Automation, Shanghai Jiaotong University, Shanghai 200240, China.
        Email: {\tt\small yinxiang@sjtu.edu.cn}}
\thanks{\quad Ziyou Song is with the Department of Electrical Engineering and Computer Science, University of Michigan, Ann Arbor, MI 48109, USA.
        Email: {\tt\small ziyou@umich.edu}}
        }%

\IEEEoverridecommandlockouts
\begin{document}

\maketitle
\begin{abstract}
Microscopic traffic models describe how cars interact with their neighbors in an uninterrupted traffic flow and are frequently used for reference in advanced vehicle control design. In this paper, we propose a novel mechanical system inspired microscopic traffic model using a mass-spring-damper-clutch system. This model naturally captures the ego vehicle's resistance to large relative speed and deviation from a (driver and speed dependent) desired relative distance when following the lead vehicle. Comparing to existing car following (CF) models, this model offers physically interpretable insights on the underlying CF dynamics, and is able to characterize the impact of the ego vehicle on the lead vehicle, which is neglected in existing CF models. Thanks to the nonlinear wave propagation analysis techniques for mechanical systems, the proposed model therefore has great scalability so that multiple mass-spring-damper-clutch system can be chained to study the macroscopic traffic flow. We investigate the stability of the proposed model on the system parameters and the time delay using spectral element method. We also develop a parallel recursive least square with inverse QR decomposition (PRLS-IQR) algorithm to identify the model parameters online. These real-time estimated parameters can be used to predict the driving trajectory that can be incorporated in advanced vehicle longitudinal control systems for improved safety and fuel efficiency. The PRLS-IQR is computationally efficient and numerically stable so it is suitable for online implementation. The traffic model and the parameter identification algorithm are validated on both simulations and naturalistic driving data from multiple drivers. Promising performance is demonstrated.
\end{abstract}

\section{Introduction}
Traffic congestion has been one of the most prevalent and stubborn challenges in urban areas for decades, causing a spectrum of issues including wasted time and economic loss \cite{inrix}, elevated driver stress and frustration \cite{driving_stress}, and increased air pollution \cite{pollution}. It is estimated that in 2017, traffic congestion costs U.S. more than \$300 billion and drivers in big cities spent more than 100 hours in congestion \cite{inrix}. To alleviate traffic congestion, various traffic control technologies have been proposed, including variable speed limits \cite{variable_speed1,variable_speed2}, dynamic traffic light control \cite{traffic_light1,traffic_light2}, and ramp metering \cite{metering1,metering2}. It is worth noting that these technologies all require accurate real-time traffic estimation and prediction. It is therefore  of critical importance to have good understanding of the traffic flow to enable those traffic control systems.

As such, numerous traffic models have been proposed to investigate traffic characteristics and flow evolution. The traffic models are generally grouped into two categories, macroscopic and microscopic. Macroscopic models are concerned with the macroscopic traffic flow characteristics such as traffic density, average speed, and traffic volume \cite{Macro}. These traffic models are inspired by continuum fluid flow theories and under different assumptions, they are further classified as kinematic models \cite{Kinematic1,Kinematic2,Kinematic3}, dynamic models \cite{Dynamic1,Dynamic2}, and lattice hydrodynamic models \cite{Hydro1,Hydro2}. On the other hand, microscopic traffic models are concerned with individual vehicles and study the local vehicle interactions in terms of speed, relative distance, and acceleration. Microscopic models can be further categorized as cellular automata (CA) models and car-following (CF) models where CA models are based on stochastic discrete event system with the ability to characterize the lane change behaviors \cite{CA1,CA2} while CF models study the ego vehicle's interaction with its preceding vehicle in a single lane \cite{1950,GHR}. CF models have great implications to the design of driving assistant systems such as adaptive cruise control \cite{review} and is the focus of this paper.

The development of CF models can date back to the 1950s \cite{1950}. Among the many CF models, the arguably most well-known model is the Gazis-Herman-Rothery (GHR) model, which was developed by the General Motors research lab in the late 1950s \cite{GHR}. The model is based on the hypothesis that the acceleration of the ego vehicle is proportional to the relative speed and inversely proportional to the relative distance, assessed at time $\tau$ earlier with $\tau$ being the delay due to reaction time. Parameters including the orders of the speed term and relative distance term, as well as a gain, were calibrated using data from wire-linked vehicles. Since then, many variants of GHR models have been developed, proposing different combinations of ``optimal'' parameters on various sets of experimental data \cite{GHR1,GHR2,GHR3}. Another class of widely-used CF models are the optimal velocity models (also referred to as Helly model), which considers a speed and/or acceleration dependent desired spacing and explicitly incorporates an error term \cite{Helly}. Several variants have also been proposed and calibrated on different experimental datasets \cite{Helly1,Helly2}. Besides the above models, some other types of models are also proposed, including collision avoidance models \cite{CA1,CA2}, psychophysical models \cite{AP1,AP2}, and fuzzy logic-based models \cite{FL}. A comprehensive review of the CF models can be referred to \cite{review}. Despite the many afore-mentioned CF models, the available relationships are still not rigorously understood and proven \cite{review}.

In this paper, we propose a new microscopic CF model, inspired by the mechanical mass-spring-damper-clutch system. There are natural similarities between the CF dynamics and the mass-spring-damper-clutch system: 1) the ego vehicle tends to accelerate when the relative distance to the lead vehicle is too large and tends to decelerate when the relative distance is too small, which resembles a mechanical spring between two masses; 2) the ego vehicle tends to follow a similar speed as the lead vehicle, resisting large speed difference. This phenomenon resembles a mechanical damper between two masses; 3) drivers tend to have delayed actions due to reaction time, which resembles a mechanical clutch whose engagement induces delays. Therefore, we propose a mass-spring-damper-clutch system to model the CF dynamics. In \cite{spring}, a mass-spring system is proposed, which is oversimplified and neglects the delayed reaction and resistance to relative speed. With the mass-spring-damper-clutch model, we further conduct stability analysis on the time delays and the related parameters using spectral element method \cite{Khasawneh2013}.

Real-time driving prediction has shown to be critical to improve fuel efficiency and road safety in advanced driving assistant systems \cite{jin}. In this study, we develop a parallel recursive  least squares with inverse QR decomposition to identify the model parameters in real-time. The algorithm is very computationally efficient and numerically stable that is suitable for the use of real-time prediction \cite{IQR}. We validate the parameter identification framework in both simulations and naturalistic driving data of three drivers. Promising performance is demonstrated.

The contributions of this paper include the following. First of all, we develop a novel mechanical system inspired mass-spring-damper-clutch system to model the CF dynamics. The new model incorporates the impacts of the ego vehicle on the lead vehicle and can be extended to estimate and predict macroscopic traffic flow with wave propagation techniques on chained mass-spring-damper systems. Secondly, we perform stability analysis on the proposed model using the spectral element method to determine the system parameter set that retains stability under different time delays. Last but not least, we develop a parallel recursive least square with inverse QR decomposition to estimate the model parameters in real time with great computational efficiency and numerical stability. Promising results are demonstrated both in simulations and on naturalistic driving data.

The remainder of this paper is organized as follows. In Section~\ref{sec:2}, we present our mechanical system inspired CF model, followed by the stability analysis of the model in Section~\ref{sec:3}. In Section~\ref{sec:4}, we present an online parameter identification algorithm using parallel recursive least squares with inverse QR decomposition. The validation of the parameter identification framework is presented in Section~\ref{sec:5}, both in simulation and on experimental data. Finally, conclusion remarks are drawn in Section~\ref{sec:6}.

\section{Mechanical System Inspired CF Model}\label{sec:2}
\begin{figure}[tp]
	\begin{subfigure}[b]{0.48\textwidth}
		\includegraphics[width=0.95\textwidth]{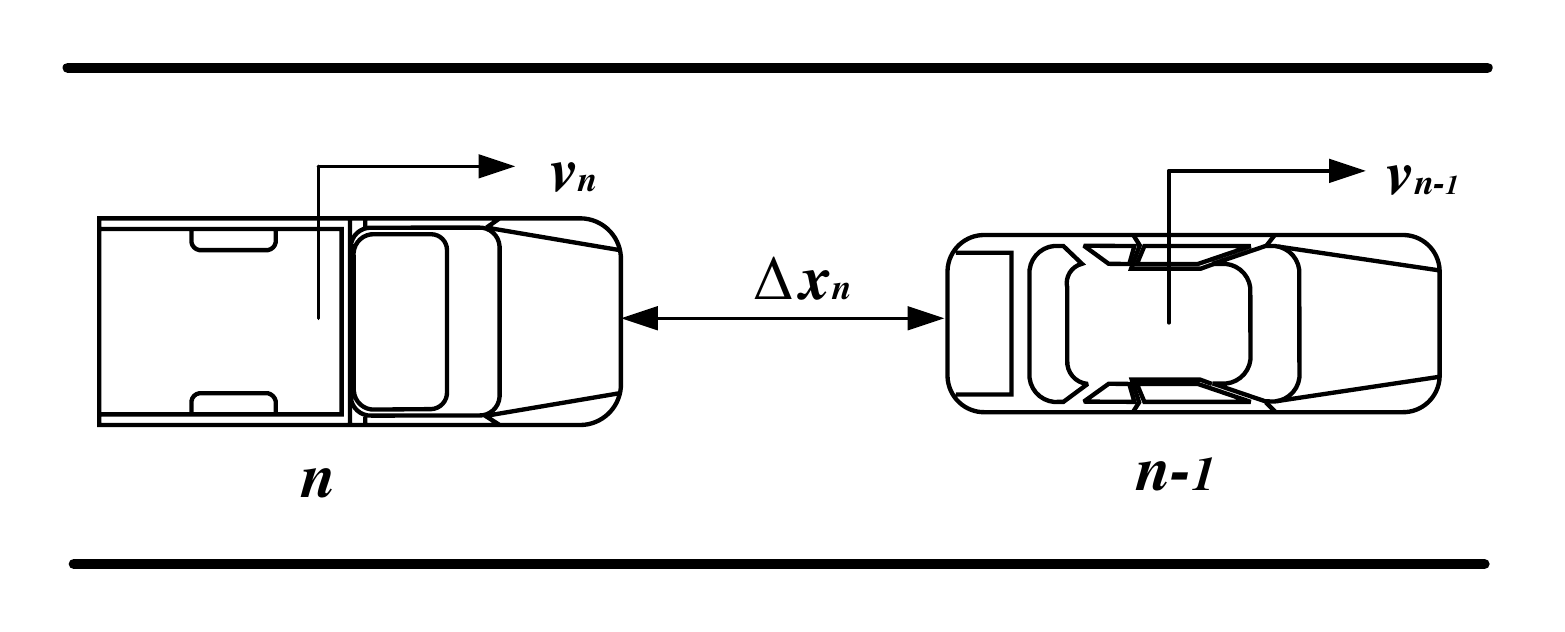}
		\caption{Schematic diagram for car-following dynamics.}\label{fig:follow}
	\end{subfigure}
	\begin{subfigure}[b]{0.48\textwidth}
		\includegraphics[width=0.95\textwidth]{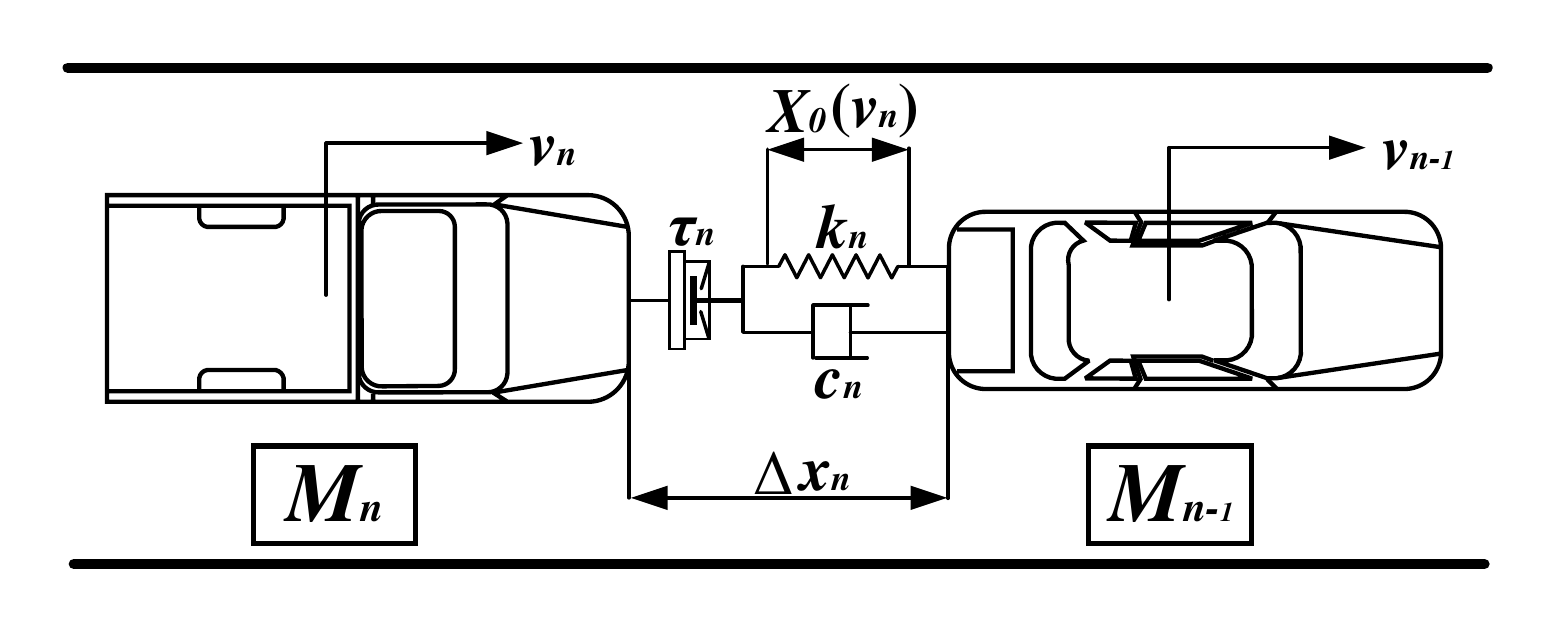}
		\caption{A mechanical mass-spring-damper-clutch system.}\label{fig:mass_spring_damper}
	\end{subfigure}
	\caption{A car following model using a mass-spring-damper-clutch system.}
\end{figure}

The car-following dynamics is illustrated in Figure~\ref{fig:follow}, where vehicle $n$ follows vehicle $n-1$ in a single lane. By convention, we name vehicle $n-1$ the lead vehicle and vehicle $n$ the follow/ego vehicle. The speeds of the ego vehicle and the lead vehicle are $v_n$ and $v_{n-1}$, respectively. The relative distance (or range) between the two vehicles is denoted as $\Delta x_n$. A car following model is characterized in terms of the ego vehicle's acceleration as a function of relative distance, vehicle speed, and relative vehicle speed \cite{GHR}:
\begin{equation}\label{equ:car_follow}
  a_n(t) = f(\Delta x(t-\tau_n), \Delta v_n(t-\tau_n), v_n(t-\tau_n)),
\end{equation}
where $a_n$ is the acceleration of the ego vehicle (vehicle $n$), $\tau_n$ is the delay due to driver reaction time and vehicle response time of the ego vehicle , and $\Delta v_n=v_{n-1}-v_n$ is the relative speed between the lead vehicle and the ego vehicle. Various mathematical models have been proposed to characterize the relationship \cite{GHR,GHR1,Helly,Helly2}. However, these models are mainly based on data regression and lack insights on the system dynamics \cite{review}.

\begin{figure}[!h]
\centering
 \includegraphics[width=0.45\textwidth]{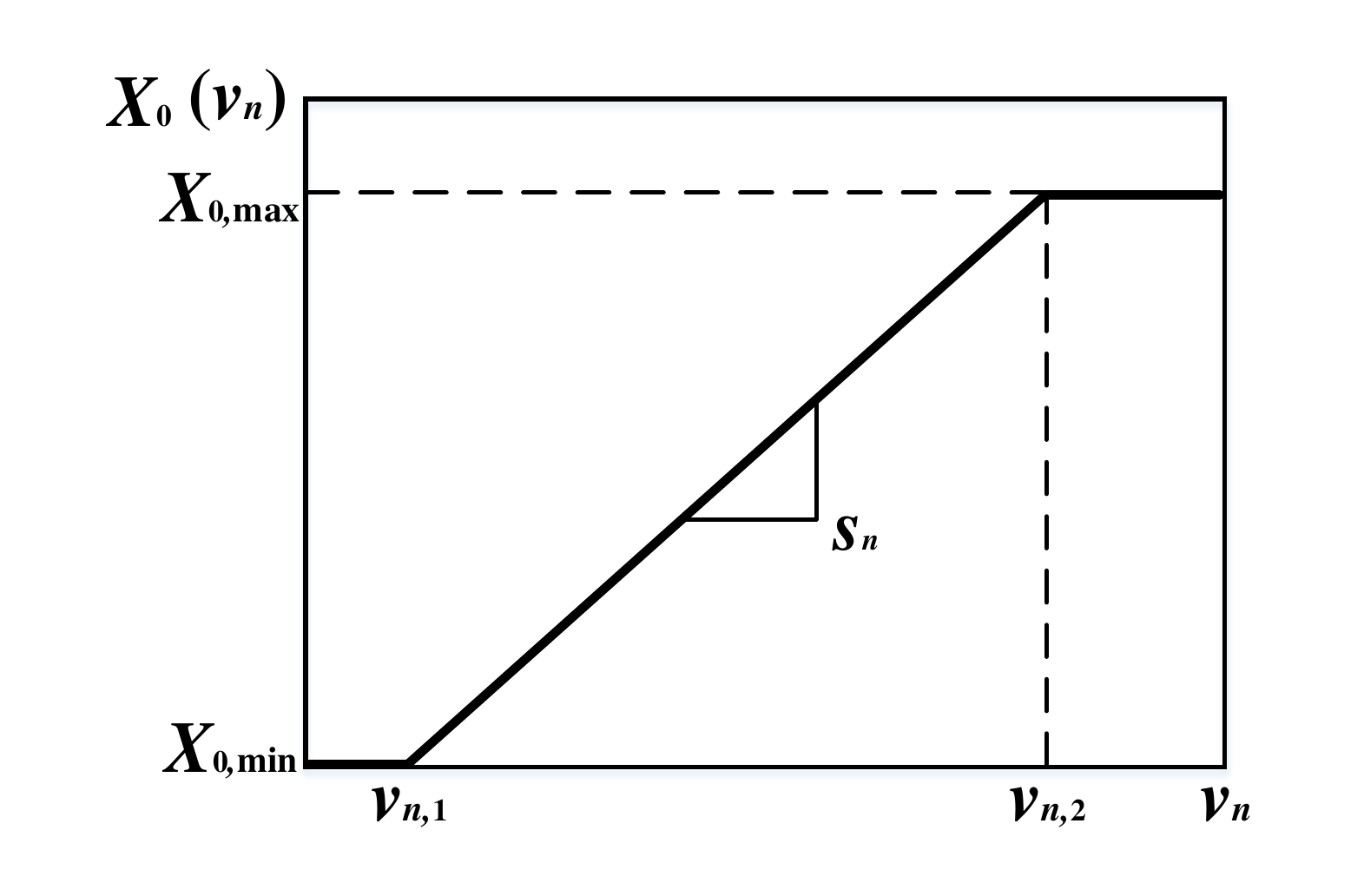}
      \caption{Speed-dependent spring relaxation length.}
      \label{fig:X0}
\end{figure}

In this paper, we propose a new mechanical system inspired CF model as shown in Figure~\ref{fig:mass_spring_damper}, where vehicle $n$ and vehicle $n-1$ are, respectively, represented as rigid bodies with masses $M_n$ and $M_{n-1}$. The two masses are connected with a spring with stiffness $k_n$, a damper with damping coefficient $c_n$, and a clutch that induces time delay $\tau_n$. Note from the observation that drivers have different desired relative distances at different vehicle speeds \cite{Helly,Helly1}, the spring in the model has the following speed-dependent relaxation length $X_0(v_n)$, which is illustrated in Figure~\ref{fig:X0}.
\begin{equation}\label{equ:relaxation}
  X_0(v_n) = \begin{cases}
               X_{0,\min}, & \mbox{if } v_n<v_{n,1} \\
               s_n\cdot v_n, & \mbox{if } v_{n,1}\leq v_n\leq v_{n,2} \\
               X_{0,\max}, & \mbox{otherwise},
             \end{cases}
\end{equation}
where $v_{n,1}$ and $v_{n,2}$ represent the lower and upper threshold points, respectively; and the slope is denoted as $s_n$.

The mass-spring-damper-clutch system naturally characterizes human driving when following a vehicle. First, drivers tend to resist large relative speed (positive or negative), which is captured by the damper that exerts forces responding to relative speed between the two vehicles. Second, drivers tend to follow a desired speed-dependent distance from the lead vehicle, which is captured by the spring with a speed-dependent relaxation length that exerts forces responding to deviations from the relaxation equilibrium. Third, the delay due to driver reaction time and vehicle response time is captured by the clutch which induces a time delay for engage and disengage.

\begin{remark}
\textit{Compared to existing CF models \cite{GHR,GHR1,GHR2,Helly,Helly1,Helly2}, the proposed model offers several advantages. First, unlike existing models that are mainly derived from data regression, this mechanical system inspired model provides interpretable physical insights on the CF dynamics. Second, the proposed model can characterize the impact of the ego vehicle on the lead vehicle, i.e., the lead vehicle tends to accelerate (if not changing lane) if the ego vehicle stays too close. This phenomenon is neglected in existing models and therefore it will cause issues when chaining CF models to represent macroscopic traffic. Third, thanks to the wave propagation techniques in mechanical mass-spring systems, the proposed model has good scalability when the mass-spring-damper-clutch systems are chained together to model macroscopic traffic flow, e.g., study the impact of shock waves.}
\end{remark}

Based on the mass-spring-damper-clutch model and the Newton's law, the equations of motion of the system can be written as:
\begin{equation}\label{equ:eom1}
\begin{aligned}
\Delta \dot{x}_n(t) &= v_{n-1}(t)-v_n(t),\\
M_{n}\dot{v}_n(t) &= k_n\left[\Delta x_n(t-\tau)-X_0(v_n(t-\tau))\right]\\
&\qquad\qquad+c_n\Delta v_n(t-\tau).
\end{aligned}
\end{equation}

In a normal highway car-following case, i.e., $v_{n,1}\leq v_n \leq v_{n,2}$, the second equation in (\ref{equ:eom1}) becomes
\begin{equation}\label{equ:eom2}
 \dot{v}_n(t) = k_n/M_{n}(\Delta x_n(t-\tau)-sv_n(t-\tau))+c_n/M_n\Delta v_n(t-\tau).
\end{equation}
Defining $x_1=\Delta x_n$, $x_2=\Delta v_n$, and $u=v_{n-1}$, (\ref{equ:eom1}) can be written as
\begin{equation}\label{equ:eom3}
\begin{aligned}
\dot{x}_1(t) &= u(t)-x_2(t),\\
\dot{x}_2(t) &= k_n/M_n\left[x_1(t-\tau)-sx_2(t-\tau)\right]\\
&\qquad\qquad+c_n/M_n(u(t-\tau)-x_2(t-\tau)).
\end{aligned}
\end{equation}

In the following section, we investigate the stability of the model on the system parameters and time delay.

\section{Stability Analysis}\label{sec:3}
In this section, we perform stability analysis of the proposed CF model in Section~\ref{sec:2}. From (\ref{equ:eom3}) and introducing the new variables $\alpha=k_n/M_n$ and $\beta=c_n/M_n$, and setting the lead vehicle's speed to be constant according to $u(t)=u(t-\tau)=u$,
equation \eqref{equ:eom3} can then be written as
\begin{equation}
\begin{aligned}
&\begin{bmatrix}
\dot{x}_1 \\
\dot{x}_2
\end{bmatrix}
=
\begin{bmatrix}
0  & -1 \\
0  & 0
\end{bmatrix}
\begin{bmatrix}
x_1(t) \\
x_2(t)
\end{bmatrix} +
\begin{bmatrix}
0       & 0\\
\alpha  & -(s \alpha +\beta)
\end{bmatrix}
\begin{bmatrix}
x_1(t-\tau) \\
x_2(t-\tau)
\end{bmatrix}\\
& +
\begin{bmatrix}
u \\
\beta\, u
\end{bmatrix},
\quad
\rm{ or } \quad
\dot{\mathbf{x}} = \mathbf{A} \mathbf{x}(t) + \mathbf{B} \mathbf{x}(t-\tau) + \mathbf{f}(t),
\label{eq:stspDDE}
\end{aligned}
\end{equation}
where $A$, $B$, and $f$ are corresponding matrices. Equation \eqref{eq:stspDDE} is a Delay Differential Equation (DDE) with a constant point delay $\tau$.
The state-space for these equations is typically taken as the space of continuous functions.
Consequently, due to the infinite dimensional nature of this state-space the stability analysis of Eq.~\eqref{eq:stspDDE} is more difficult than its delay-free counterpart.
Nonetheless, there are several methods available for the stability analysis of this problem including the semi-discretization method \cite{Insperger2004}, Chebyshev polynomials \cite{Butcher2004}, and the spectral element method (SEM) \cite{Khasawneh2013}.
In this paper we will use the SEM approach due to its flexibility and efficiency \cite{Tweten2012}.

The main idea of the SEM is to discretize the state space of Eq.~\eqref{eq:stspDDE} and then construct a dynamic map over one period where the length of this period for autonomous systems is typically taken to be the length of the time delay $\tau$.
The eigenvalues of the resulting matrix that describes this dynamic map must be within the unit disc of the complex plan in order for the corresponding DDE to be stable.
While convergence in the SEM can be obtained either by using multiple temporal elements ($h$-refinement) or by increasing the order of the interpolating polynomial ($p$-refinement), in this study we use one temporal element and only increase the order of the polynomials to achieve convergence.

Since the vector $\mathbf{f}$ in Eq.~\eqref{eq:stspDDE} only affects the steady state solution but does not affect the stability analysis, we drop it from the subsequent discussion.
Let $\mathbf{T}=\{t_i\}_{i=1}^{n+1}$ be a set of $n+1$ distinct temporal mesh points on $[0, \tau]$, and let $\mathbf{c}_m$ and $\mathbf{c}_{m-1}$ be $2\times (n+1)$ vectors containing the values of the states $\mathbf{x}(t)$ and the delayed states $\mathbf{x}(t-\tau)$, respectively, evaluated on $\mathbf{T}$.
We choose a barycentric Lagrange interpolation \cite{Butcher2004} to represent the states according to
\begin{equation}
\mathbf{x}(t) = \boldsymbol{\Phi} \, \mathbf{c}_m, \text{ and } \mathbf{x}(t-\tau) = \boldsymbol{\Phi} \, \mathbf{c}_{m-1},
\end{equation}
 where $\boldsymbol{\Phi}=\boldsymbol{\Phi}(t)=(\boldsymbol{\phi}(t) \otimes \mathbf{I})$, and $\boldsymbol{\phi}(t)$ is the vector of barycentric Lagrange interpolating polynomials $\boldsymbol{\phi}(t) = [L_1(t), L_2(t), \ldots, L_{n+1}(t)]$, $\mathbf{I}$ is the $2 \times 2$ identity matrix, while $\otimes$ is the Kronecker product.
We now substitute the state approximations into the DDE to obtain
\begin{equation}
(\dot{\boldsymbol{\Phi}} - \mathbf{A} \boldsymbol{\Phi}) \mathbf{c}_m = \mathbf{B} \boldsymbol{\Phi} \mathbf{c}_{m-1} + \boldsymbol{\epsilon},
\end{equation}
where $\boldsymbol{\epsilon}$ is the vector of approximation errors.
Let $\boldsymbol{\psi}=[\psi_1(t), \psi_2(t), \ldots, \psi_{n+1}(t)]$ be a vector of linearly independent test functions.
This set of functions is then used in a Galerkin approach where the errors are required to be perpendicular to the space spanned by the set $\boldsymbol{\psi}$.
The result is the $2(n+1) \times 2(n+1)$ system of equations
\begin{equation}
\label{eq:dynMap1}
\left(\int\limits_{0}^{\tau}{\boldsymbol{\Psi}  \left(\dot{\boldsymbol{\Phi}}-\mathbf{A}\, \boldsymbol{\Phi}\right) dt}\right) \mathbf{c}_m =
	\left(\int\limits_{0}^{\tau}{\boldsymbol{\Psi} \, \mathbf{B}\, \boldsymbol{\Phi} \, dt}\right) \mathbf{c}_{m-1},
\end{equation}
where $\boldsymbol{\Psi}=(\boldsymbol{\psi} \otimes \mathbf{I})^T$.
Note that the integrals in Eq.~\eqref{eq:dynMap1} are often difficult to evaluate analytically which necessitates using numerical integration as described in \cite{Khasawneh2013}.
Equation \eqref{eq:dynMap1} can then be used to construct a dynamic map $\boldsymbol{\Gamma}$ according to
\begin{equation}
\label{eq:dynMap2}
\begin{aligned}
 \mathbf{c}_m& =
	\left(\int\limits_{0}^{\tau}{\boldsymbol{\Psi}  \left(\dot{\boldsymbol{\Phi}}-\mathbf{A}\, \boldsymbol{\Phi}\right) dt}\right)^{-1}\left(\int\limits_{0}^{\tau}{\boldsymbol{\Psi} \, \mathbf{B}\, \boldsymbol{\Phi} \, dt}\right) \mathbf{c}_{m-1}\\
& = \boldsymbol{\Gamma} \mathbf{c}_{m-1}.
\end{aligned}
\end{equation}
In order to ascertain the stability of Eq.~\eqref{eq:stspDDE}, we examine the eigenvalues of $\boldsymbol{\Gamma}$: if all the eigenvalues are within a modulus of less than one in the complex plane then the system is asymptotically stable.

In this paper we used the SEM with a $100 \times 100$ grid in the $(\alpha, \beta)$ plane where $\alpha \in [0.01, 2]$, $\beta \in [0.01, 8]$, while 6 equally spaced values of $\tau$ were considered in the range $\tau \in [0.2, 2]$.
The temporal mesh used consisted of $21$ Legendre-Gauss-Lobatto points which correspond to an interpolating polynomial of order $20$, while the trial functions were the shifted Legendre polynomials.
Increasing the order of the interpolating polynomial beyond 20 did not change the results, thus indicating the convergence of the solution.

\begin{figure}
\begin{minipage}{0.5\textwidth}
\centering
\includegraphics[width=\textwidth]{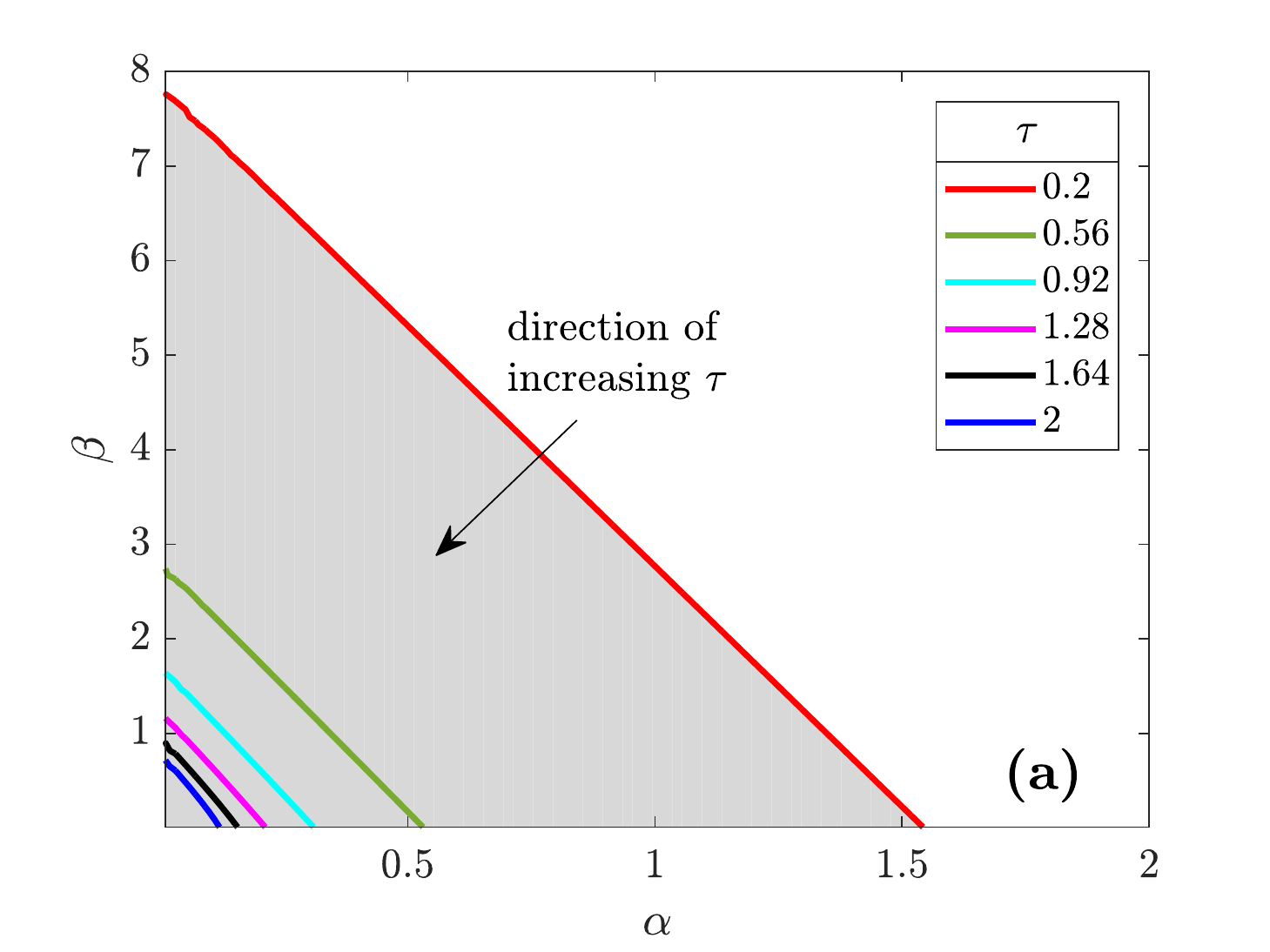}
\end{minipage}
\begin{minipage}{0.5\textwidth}
\centering
\includegraphics[width=\textwidth]{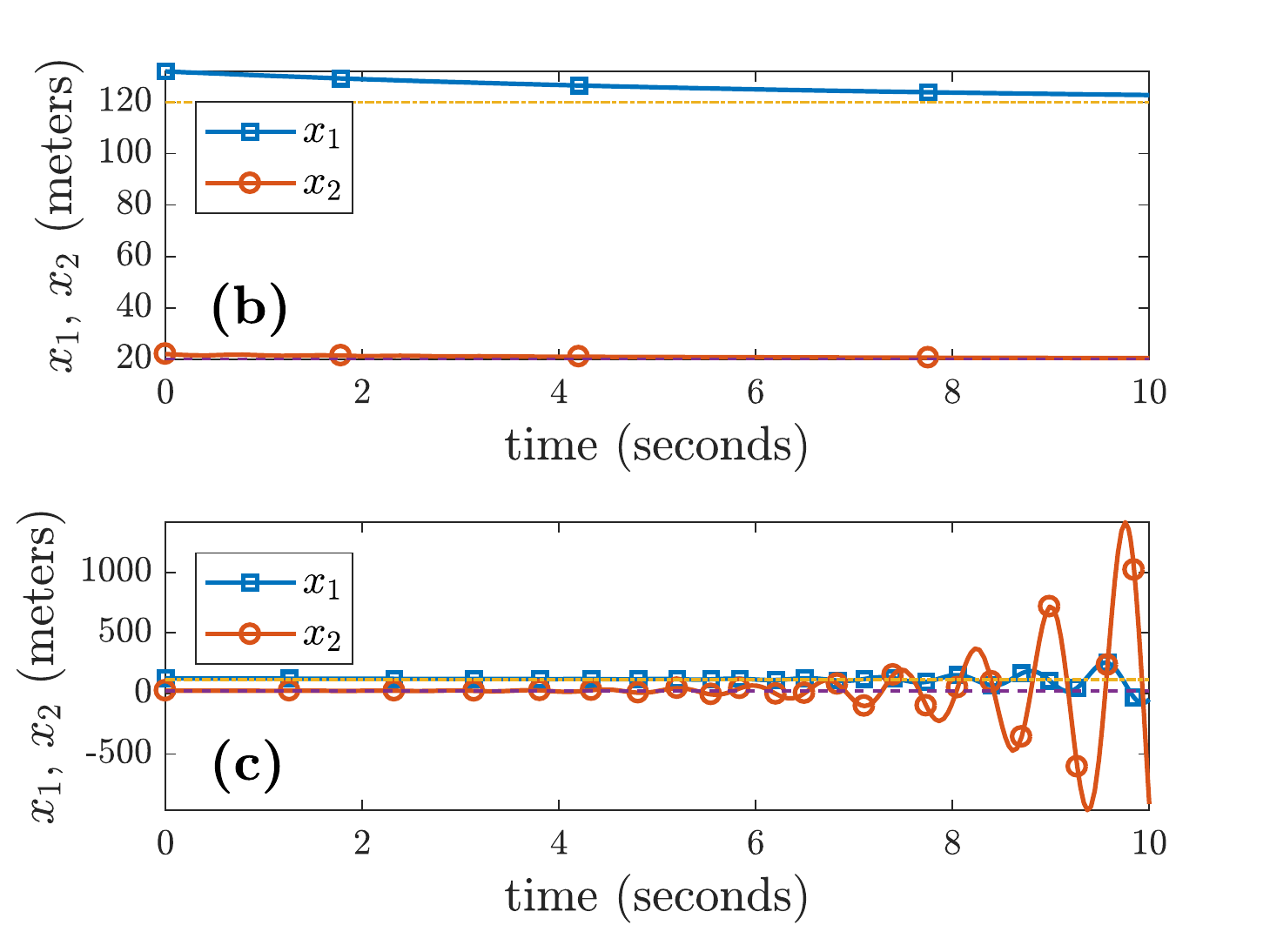}
\end{minipage}
\caption{(a) The stability diagram in the $(\alpha, \beta)$ plane for increasing values of $\tau$ as well as the simulated time series for $\tau=0.2$,
(b) ($\alpha=1$, $\beta=2$), and (c) ($\alpha=1.6$, $\beta=2$).}
\label{fig:stab_results}
\end{figure}

Figure~\ref{fig:stab_results} shows the stability diagram in the $(\alpha, \beta)$ plane for 6 different values of $\tau$.
The shaded region under each curve is stable, while the area above the curves are unstable.
It can be seen that the triangular stability regions shrink as the value of the delay parameter increases from $0.2$ to $2$.
In order to show the resulting stable and unstable system responses two points were simulated in Figs.~\ref{fig:stab_results}b and c using Matlab's dde23 function and a value of $u=20$ in Eq.~\eqref{eq:stspDDE}.
The chosen stable point is $(\alpha=1, \beta=2, \tau=0.2)$, while the selected unstable point is $(\alpha=1.6, \beta=2, \tau=0.2)$.
The history function used in the simulation was the perturbed steady state solution $\tilde{\mathbf{x}}=\mathbf{x}_{\rm stst}(1+0.10)$ where $\mathbf{x}_{\rm stst}$ is the constant steady state solution given by
\begin{equation}
\mathbf{x}_{\rm stst} = \begin{bmatrix}
x_1 \\
x_2
\end{bmatrix}_{\rm stst} =
\frac{u}{\alpha}\,\begin{bmatrix}
-1 + \alpha s + \beta\\
\alpha
\end{bmatrix}.
\end{equation}
Figure~\ref{fig:stab_results}b shows that the perturbed system goes back to the steady state solution as the system evolves.
In contrast, the perturbed, unstable system of Fig.~\ref{fig:stab_results}c deviates from the steady state solution and grows exponentially.

\section{Online Parameter Identification}\label{sec:4}
Vehicle speed prediction is essential in automated longitudinal control for improved fuel efficiency and safety \cite{CCC}. Existing studies on cooperative cruise control generally assume that the dynamics of the preceding vehicles are available \cite{CCC1,CCC2}. While this assumption is valid for fully autonomous vehicle platoon, it does not hold if human drivers are involved. Therefore, it is important to accurately predict human driver maneuvers to provide a ``comprehensive preview''. In this paper, we develop a framework to identify the parameters and predict vehicle speed changes online.

We first discretize (\ref{equ:eom2}) using the explicit Euler method with sampling time $\Delta t$, which gives
\begin{equation}\label{eom_discretized}
\begin{aligned}
\frac{v_n(k)-v_n(k-1)}{\Delta t} &= k_n/M_n(\Delta x_n(k-d)-s_nv_n(k-d))\\
&\qquad+c_n/M_n\Delta v_n(k-d),
\end{aligned}
\end{equation}
where $d=\text{round}(\tau_n/\Delta t)$ is the corresponding delayed steps.

For a specific delay $d$, define $\alpha_{n}(d)=k_n/M_n$, $\beta_{n}(d)=-(k_ns_n)/M_n$, and $\gamma_{n}(d)=c_n/M_n$, then (\ref{eom_discretized}) can be written as the following linear equation:
\begin{equation}\label{equ:linear}
\begin{aligned}
  &[\Delta x_n(k-d)\; v_n(k-d)\; \Delta v_n(k-d)]\begin{bmatrix}
                                                  \alpha_n(d) \\
                                                  \beta_n(d) \\
                                                  \gamma_n(d)
                                                \end{bmatrix}\\
                                                &\qquad\qquad = \frac{v_n(k)-v_n(k-1)}{\Delta t}.
\end{aligned}
\end{equation}
Given a time series data of $K$ steps, $K>d$, and define the parameter vector $p_n(d)=[\alpha_n(d);\beta_n(d);\gamma_n(d)]$, then the parameters can be identified by solving the following least-square problem:
\begin{equation}\label{equ:optimization}
\min_{p_n(d)}\left\|A_n(d)p_n(d)-B_n(d)\right\|^2,
\end{equation}
where $A_n(d)=\begin{bmatrix}
           \Delta x_n(0) & v_n(0) &\Delta v_n(0) \\
           \Delta x_n(1) & v_n(1) & \Delta v_n(1) \\
           \vdots &\vdots & \vdots \\
           \Delta x_n(N-d) & v_n(N-d) &\Delta v_n(N-d)
         \end{bmatrix}$ and $B_n(d)=\frac{1}{\Delta t}\cdot\begin{bmatrix}
                                    v_n(d)-v_n(d-1) \\
                                    v_n(d+1)-v_n(d) \\
                                    \cdots \\
                                    v_n(N)-v_n(N-1)
                                  \end{bmatrix}$ are the data matrices of vehicle $n$. Note that for different delay $d$s, the data matrices are different, which leads to different identified parameters. In this paper, we consider the possible range of the delay $\tau_n\in\left[\tau_{\min},\tau_{\max}\right]$. With the sampling time $\Delta t$, the range of the discrete time delay is $d\in\{d_{\min},d_{\min}+1,\cdots,d_{\max}\}$, where $d_{\min}=round(\tau_{\min}/\Delta t)$ and $d_{\max}=round(\tau_{\max}/\Delta t)$.

It is straightforward to show that the optimal solution to (\ref{equ:optimization}) is
\begin{equation}\label{equ:sol}
p_n^*(d)=(A_n^\text{T}(d)A_n(d))^{-1}A_n^\text{T}(d)B_n(d).
\end{equation}

Note that the sizes of matrices $A_n$ and $B_n$ increase as the data length grows, causing computational issues if implemented online. Therefore, recursive computation is needed. In this paper, we exploit a recursive least squares with inverse QR decomposition algorithm (RLS-IQR) for online identification, which has great numerical stability and computational efficiency \cite{IQR}. Specifically, at each time step $k$, $k>d$, the algorithm takes in the input vector $x(k)=[\Delta x_n(k-d)\; v_n(k-d)\; \Delta v_n(k-d)]$ and output $y(k)=\frac{v_n(k)-v_n(k-1)}{\Delta t}$, and then update the parameters $p_n(d)$. The details of the update is shown in Algorithm~\ref{algo:1}. Since there are multiple possible delays, we run the RLS-IQR in parallel for each $d\in\{d_{\min},d_{\min}+1,\cdots,d_{\max}\}$. To determine the best delay $d$ and the corresponding parameter $p_n^*(d)$ for prediction, we accumulate the prediction error as:
\begin{equation}\label{equ:accum_error}
J(d,k+1) = (1-\alpha)J(d,k)+\alpha |y(k+1)-x(k+1)p_n^*(d)|,
\end{equation}
with $J(d,0)=0$ for all $d\in\{d_{\min},d_{\min}+1,\cdots,d_{\max}\}$ and $\alpha\in(0,1]$ is the learning rate. To determine the best delay parameter $d$, we use the delay parameter corresponding to the minimum accumulated error:
\begin{equation}\label{equ:bestp}
  d^*(k) = \arg\min_d J(d,k).
\end{equation}
Then the parameter for prediction is chosen as $p^*(k)=p_n(d^*)$. The process is summarized in Algorithm~\ref{algo:1}.
\begin{figure}[H]
\removelatexerror
\begin{algorithm}[H]
\SetAlFnt{\small}
    \SetKwInOut{Input}{Input}
    \SetKwInOut{Output}{Output}
\caption{Parallel Recursive Least Squares with Inverse QR Decomposition}
\label{algo:1}
\SetAlgoLined
\Parameter{Accumulated error learning rate $\alpha$, RLS forgetting factor $\lambda$, inverse matrix initialization parameter $\delta$.}
\Input{$\{\Delta x_n(k),v_n(k),\Delta v_n(k)\}_{k=1}^N$.}
\Output{$\{p_n(d,k),J_n(d,k)\}_{k=1}^{N},d=d_{\min},\cdots,d_{\max}$.}
\vspace{0.2em}
\hrule
\vspace{0.2em}
{\bf initialize} $p_n(d,0) \gets \mathbf 0_{3\times1}$, $R^{-\text T}(d) \gets \delta \mathbf I_3$, $J(d,0)\gets0$, $d=d_{\min},\cdots,d_{\max}$;\\
\For{$k=1 \to N$}{
    \tcc{ \it Parallel loop for possible delays}
    \For{$d=d_{\min}\to d_{\max}$}{
        \If{$k\ge d$}{
        {\bf set} {\small{$x^\text{T}(k)\gets [\Delta x_n(k-d)\; v_n(k-d)\; \Delta v_n(k-d)], y^\text{T}(k)\gets\frac{v_n(k)-v_n(k-1)}{\Delta t}$;}}\\
        \tcc{ \it Accumulate prediction error}
        {\bf compute} $e_d(k|k-1)=y(k)-x^\text{T}(k)p_n(d,k)$;\\
        $J(d,k)\gets (1-\alpha)J(d,k-1)+\alpha |e_d(k|k-1)|$;\\

        \tcc{\it Parameter update}
        {\bf initialize} $u_{j,m}\gets 0, b_0\gets 1, 1\leq j\leq 3, \, m<j;$\\

        \For{$i=1\to 3$}{
        $a_i=\lambda^{-1/2}\sum_{j=1}^ir_{ij}(d,k-1)x_k(j)$;\\
        $b_i=\sqrt{b_{i-1}^2+a_i^2}$;\\
        $s_i=a_i/b_i;$\\
        $c_i=b_{i-1}/b_i;$\\
            \For{$j=1\to i$}{
                $r_{ij}(k)=\lambda^{-1/2}c_ir_{ij}(k-1)-s_iu_{i-1,j}$;\\
                $u_{i,j}=c_iu_{i-1,j}+\lambda^{-1/2}s_ir_{ij}(k-1)$;\\
            }

        }
        $z(k)=e(k|k-1)/b_3$;\\
        \For{$i=1\to 3$}{
        $p_n(d,k)\gets p_n(d,k-1)+z(k)u_{i,3}$;\\
        }
    }
}
}
\end{algorithm}
\end{figure}

\section{Simulation Validation}\label{sec:5}
\begin{figure}[!h]
\centering
 \includegraphics[width=0.4\textwidth]{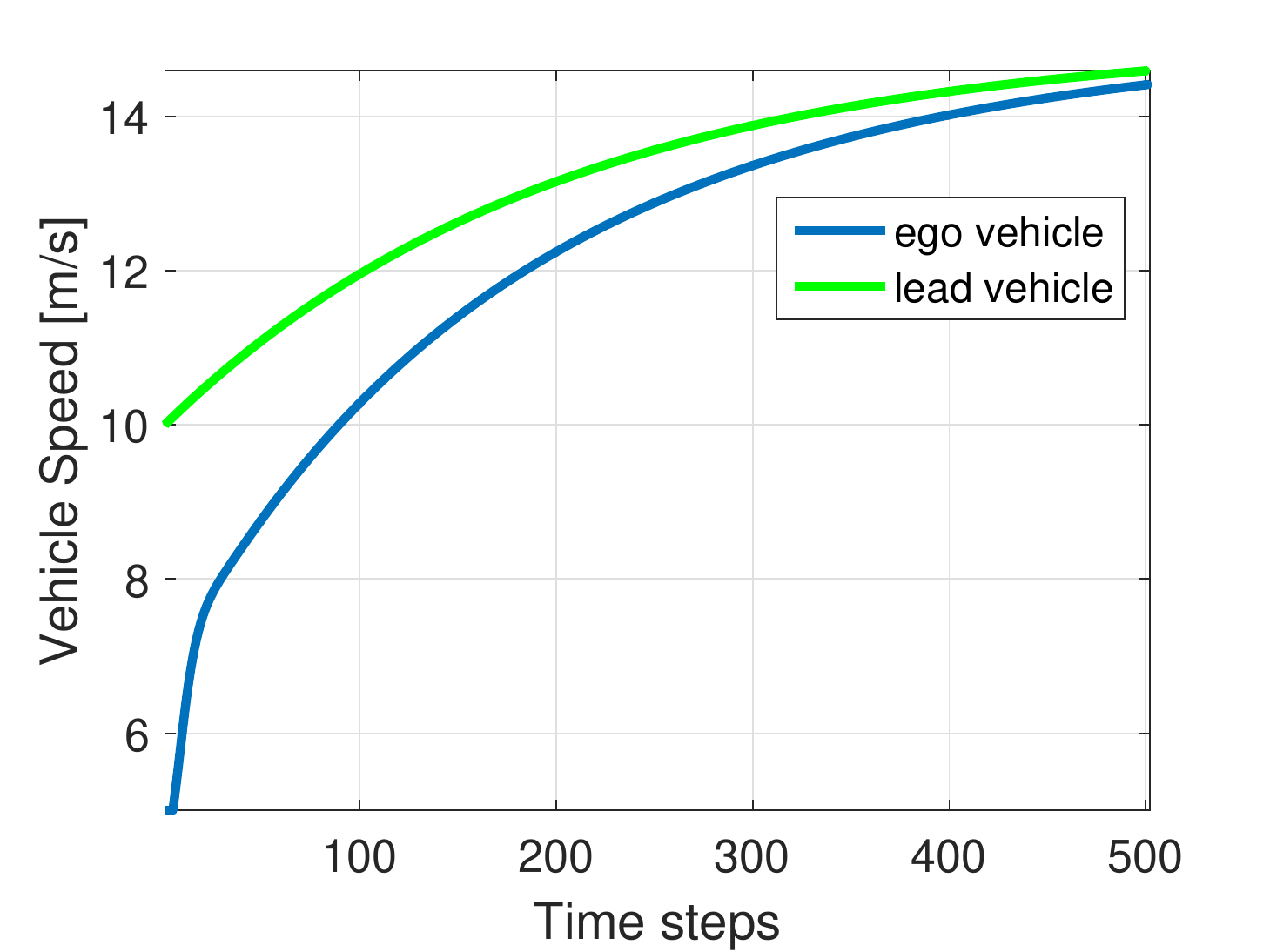}
      \caption{Simulated vehicle speeds.}
      \label{fig:speed}
\end{figure}

In this section, we perform simulation to validate the developed online parameter identification algorithm. Towards that end, we use the mass-spring-damper-clutch system with the parameters listed in Table~\ref{tab:1}. The vehicle following is simulated over a 50 seconds horizon with initial ego vehicle speed 5 $m/s$ and initial distance headway 20 $m$. The lead vehicle speed profile is set as $v_l=15-5\exp(-0.05t)$. Following the dynamics (\ref{equ:eom3}), the ego vehicle speed and the relative distance is obtained. The speeds of the lead vehicle and ego vehicle are shown in Figure~\ref{fig:speed}.

\begin{table}[!h]
\caption{Parameters for vehicle following simulation.}
\label{tab:1}
\centering
\begin{tabular}{|c|c|c|c|c|c|}
\hline
\hline
 $M_n[kg]$ &$k_n[N/m]$ &$c_n[N\cdot s/m]$ &$s_n [s]$ &$\tau_n [s]$ &$\Delta t [s]$\\
 \hline
1000 &100 &500 &5 &0.5 &0.1\\
 \hline\hline
\end{tabular}
\end{table}

\begin{figure}[!h]
\centering
 \includegraphics[width=0.45\textwidth]{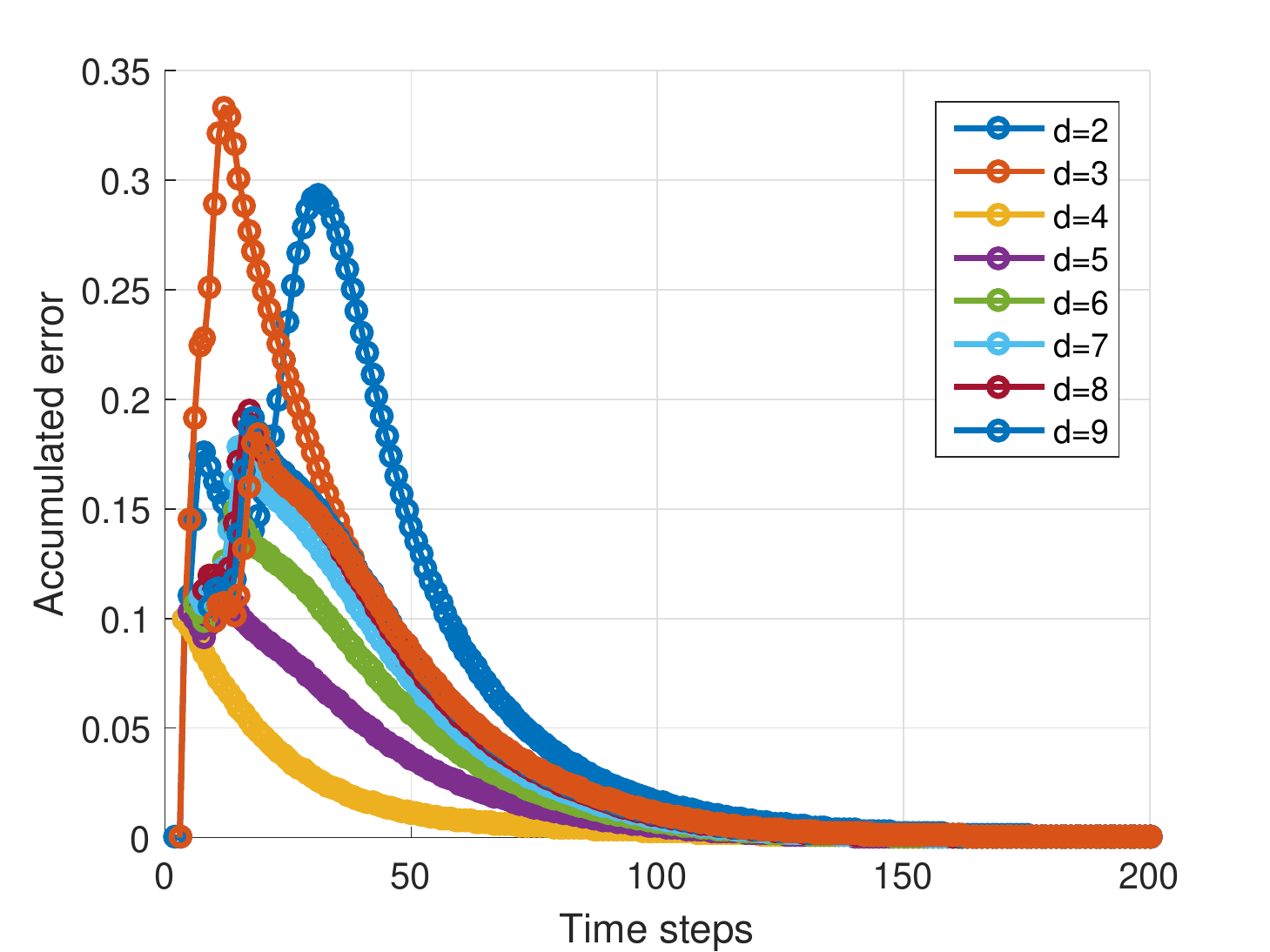}
      \caption{Accumulated prediction error for possible delays.}
      \label{fig:errors}
\end{figure}

We apply the Parallel RLS-IQR algorithm in Section~\ref{sec:4} on the simulated data. We consider the reaction time range as $\tau_{\min}=0.2[s]$ and $\tau_{\max}=1[s]$. With the sampling time $\Delta t=0.1$, the corresponding delays are $d_{\min}=2$ and $d_{\max}=10$. As a result, we run a parallel of 9 RLS-IQR for each of the possible delays. The forgetting factor $\lambda$ is chosen as 0.95 and the inverse matrix initialization parameter is chosen as 10. The accumulated prediction error with learning rate $\alpha=0.05$ for all the delays are shown in Figure~\ref{fig:errors}. It can be seen that $d=4$ gives the lowest prediction error, which matches the specified time delay in the simulation.

\begin{figure}[!h]
\centering
 \includegraphics[width=0.45\textwidth]{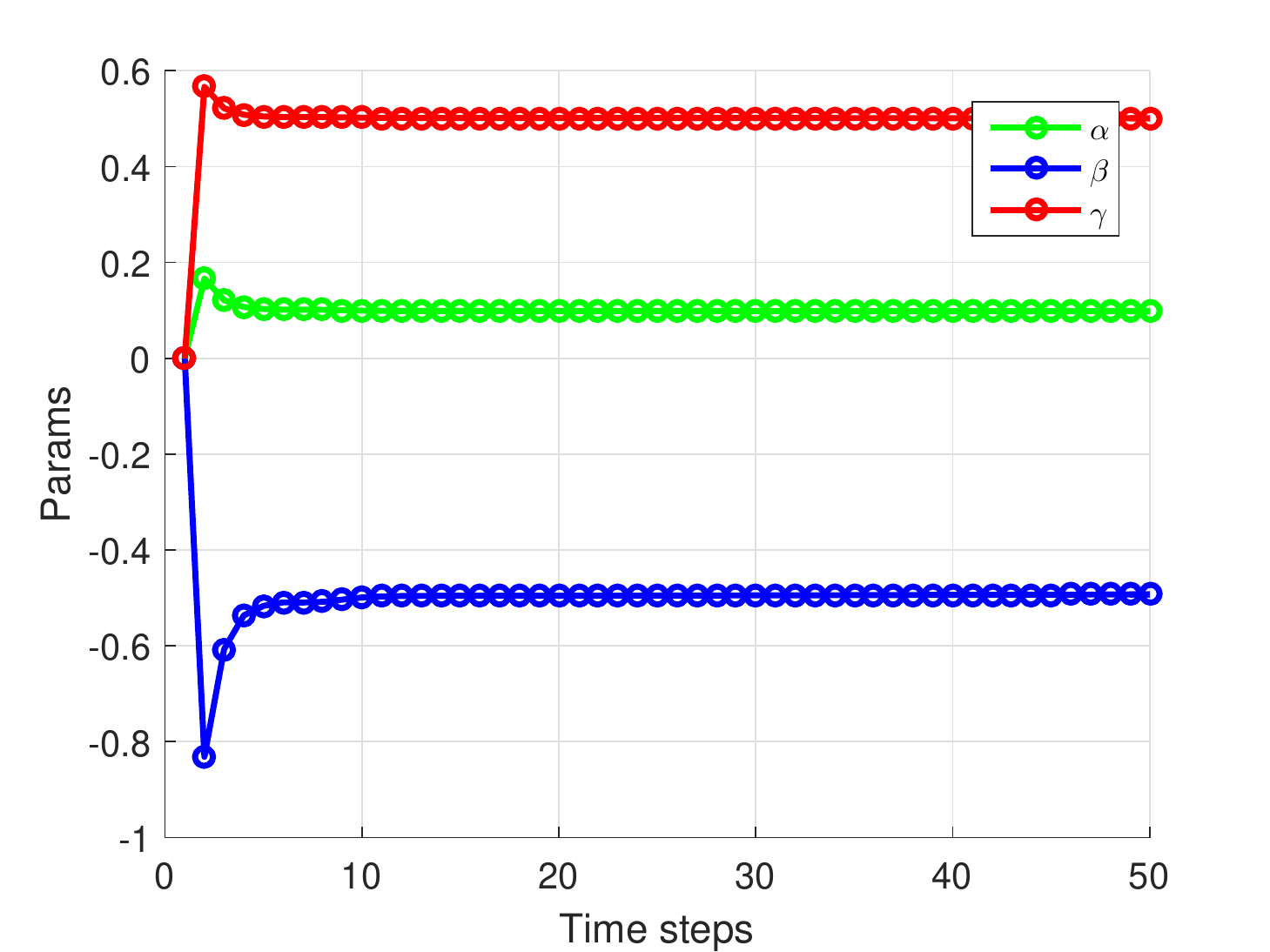}
      \caption{Online estimated parameters. It can be seen that it only takes 10 steps (1 sec) to correctly identify the parameters in simulation.}
      \label{fig:params}
\end{figure}

The online estimated parameters are shown in Figure~\ref{fig:params}. It can be seen that in the ideal simulation case, i.e., no model uncertainty and perfect measurement, it only takes 10 steps (1 second) to identify the parameters. This justifies the use of online prediction. However, in reality sensor noises exist in the measurement channel. We next investigate the sensitivity of algorithms to the measurement noises.

\begin{figure}[!h]
\centering
 \includegraphics[width=0.45\textwidth]{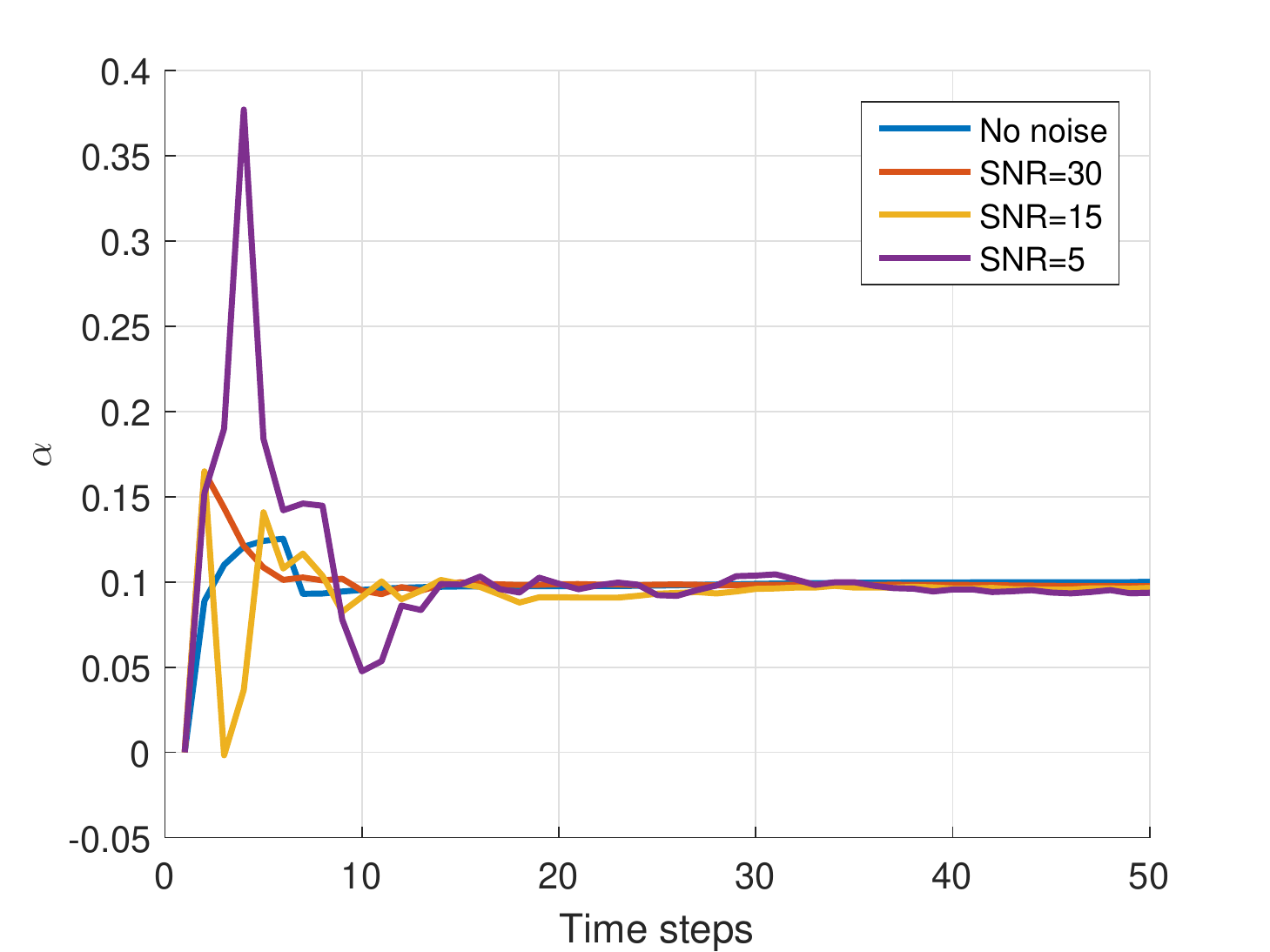}
      \caption{Estimation of $\alpha$ under different noise levels.}
      \label{fig:alpha}
\end{figure}

\begin{figure}[!h]
\centering
 \includegraphics[width=0.45\textwidth]{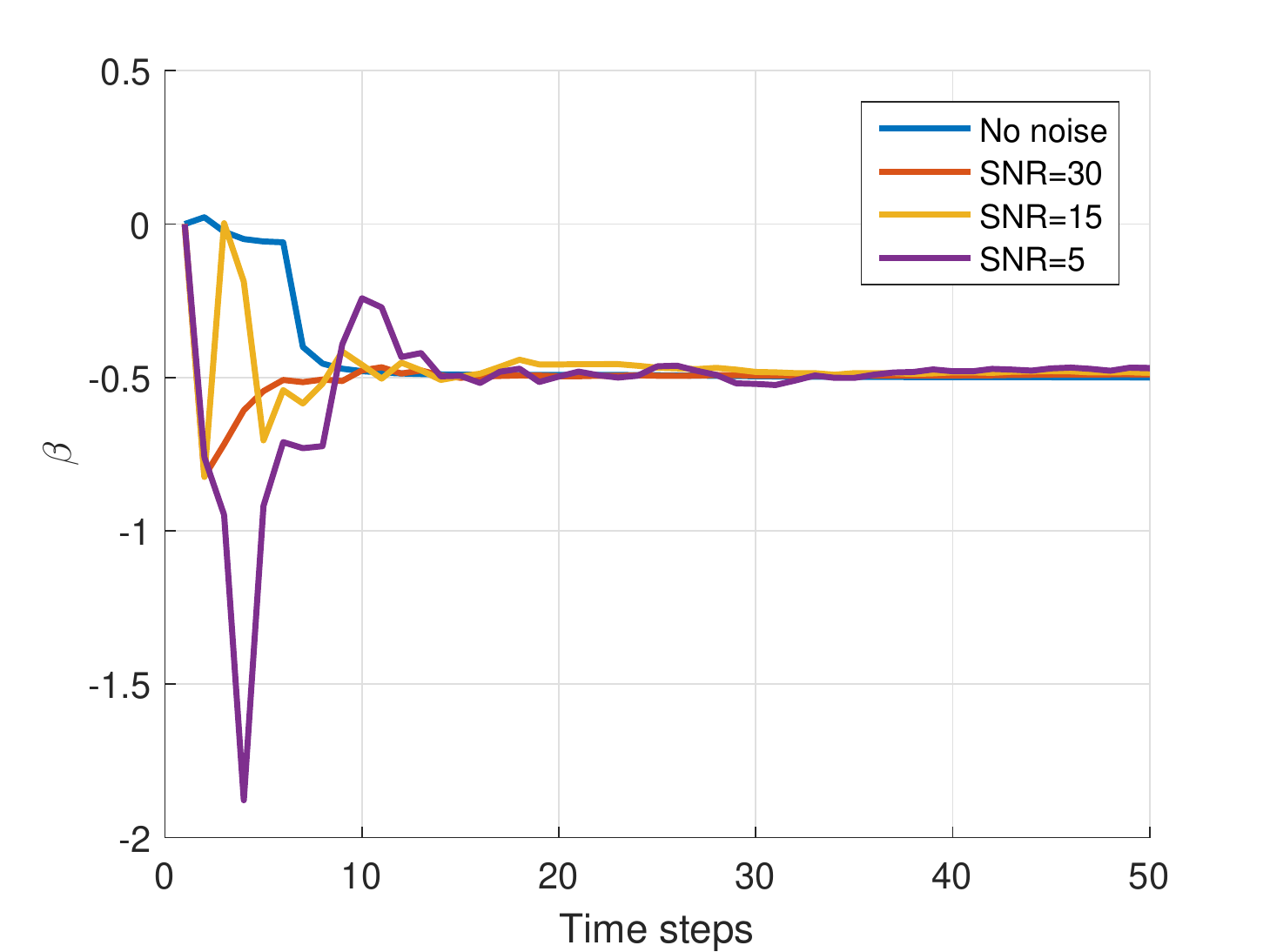}
      \caption{Estimation of $\beta$ under different noise levels.}
      \label{fig:beta}
\end{figure}

\begin{figure}[!h]
\centering
 \includegraphics[width=0.45\textwidth]{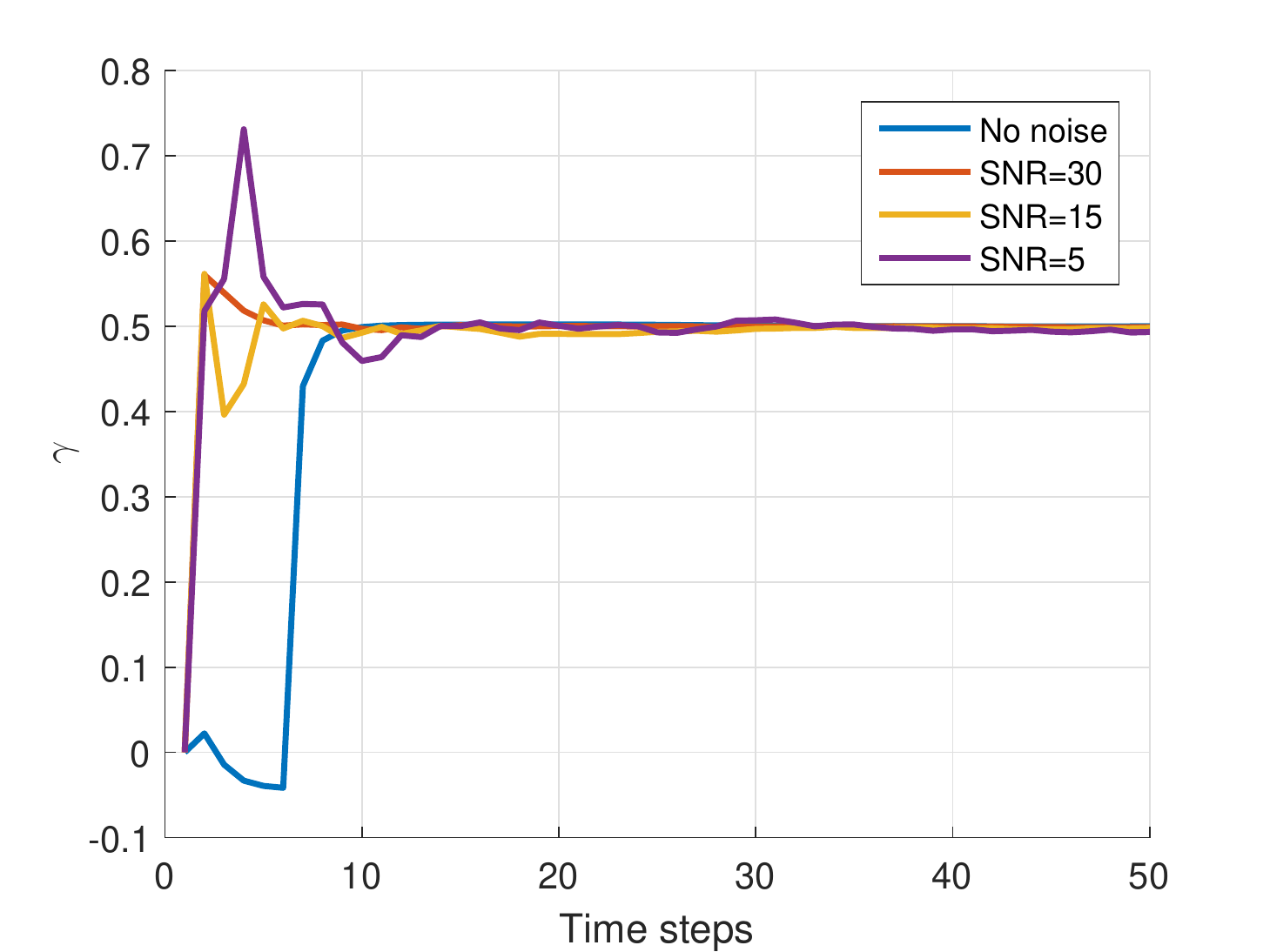}
      \caption{Estimation of $\gamma$ under different noise levels.}
      \label{fig:gamma}
\end{figure}

Towards this end, we inject measurement noises in the measurement channels. We consider 4 levels of noises with respect to signal-to-noise ration (SNR). The SNR of a discrete signal $y$ of length $N$ with noise $e$ is defined as
\begin{equation}\label{equ:SNR}
  SNR = 10 \log_{10}\frac{\sum_{k=1}^N(y(k)-e(k))^2}{\sum_{k=1}^Ne^2(k)}.
\end{equation}
In the simulation, we inject Gaussian noises to ego vehicle speed, relative distance, relative speed, and acceleration of the ego vehicle. We performed the online system identification on noise levels: no noise, SNR 30 dB, SNR 15 dB, and SNR 5 dB. The parameter estimation performance is shown in Figures~\ref{fig:alpha}-\ref{fig:gamma}. It can be seen that the algorithm is robust to noises and have fast convergence rate even under large noise levels.

\section{Model validation on Naturalistic Driving Data}

\begin{figure}[!h]
\centering
 \includegraphics[width=0.45\textwidth]{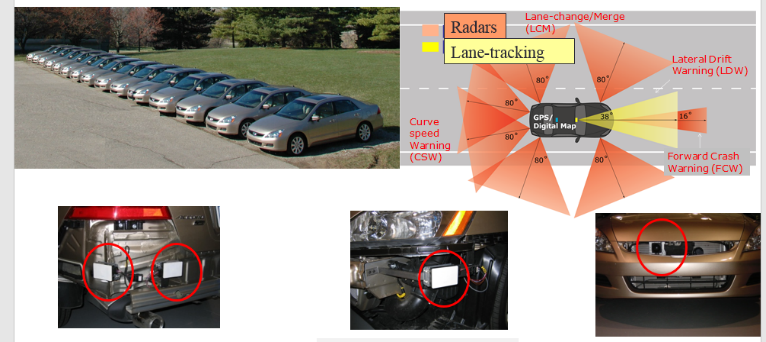}
      \caption{Vehicle fleet and instrumentation for the IVBSS program.}
      \label{fig:fleet}
\end{figure}

In this section, we validate the proposed mass-spring-damper-clutch CF model and the online parameter estimation algorithm on a naturalistic driving dataset from the Integrated Vehicle based Safety System (IVBSS) program \cite{IVBSS}. The main objective of the IVBSS program was to investigate the effectiveness of driving assistant systems such as Lane Departure Warning, Curve Speed Warning, and Forward Crash Warning. A diverse group of 108 drivers participated in the program with balanced age and gender. The participants drove the experimental vehicles for their personal use for about six weeks. The experimental vehicles are equipped with data collection instruments to record driving data including vehicle speed and acceleration, as well as the relative distance and relative speed to the leading vehicle that are estimated using Mobileye \cite{Eye}. Data including vehicle speed, acceleration, relative distance and relative distance are recorded every 0.1 second (10 Hz). The vehicle fleet and some instrumentations are illustrated in Figure~\ref{fig:fleet}.

\begin{figure*}[!ht]
\centering
 \includegraphics[width=0.95\textwidth]{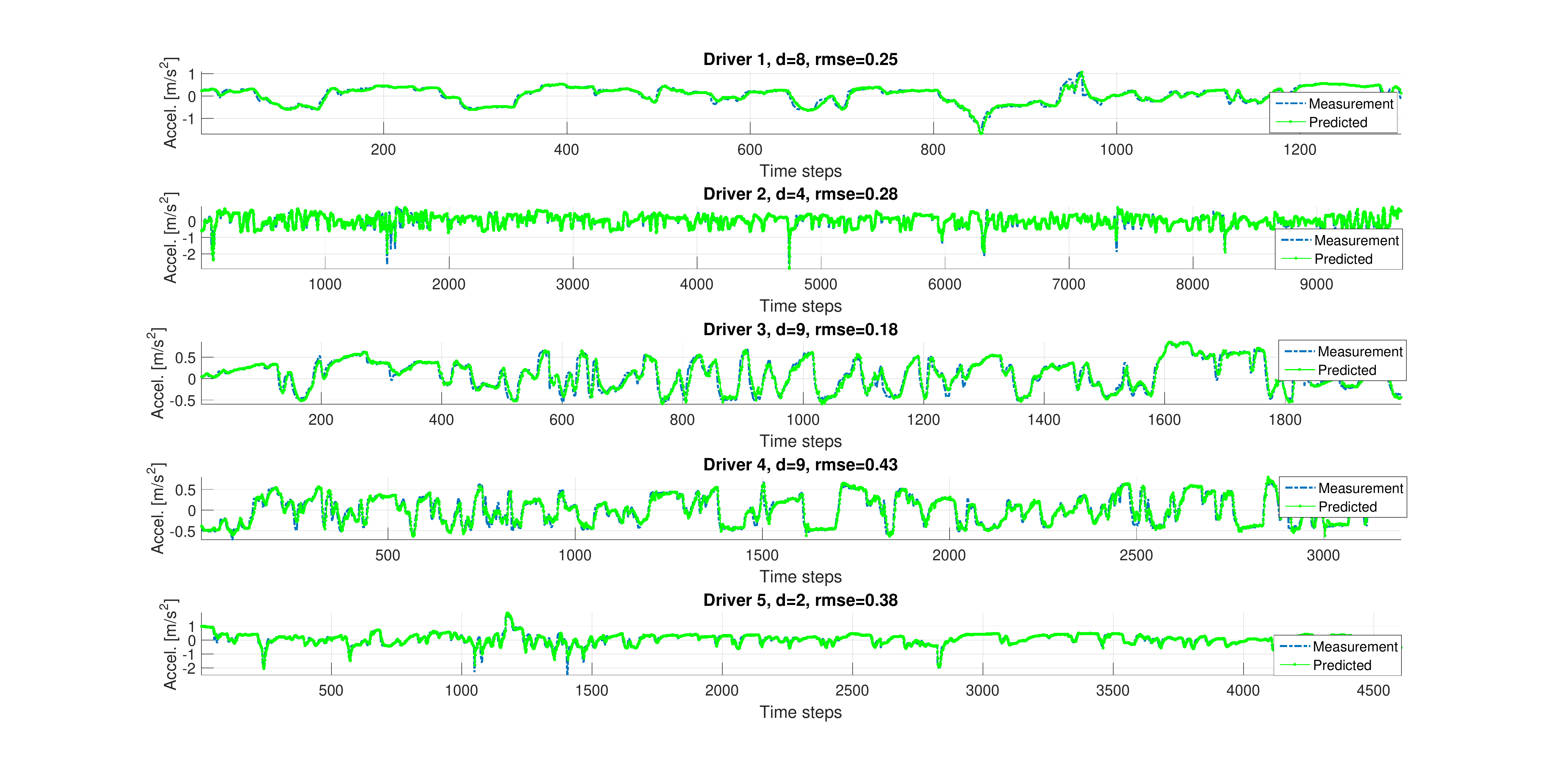}
      \caption{Sample vehicle acceleration and online predictions for five drivers. The prediction is based on best approximated delays, i.e., the lowest accumulated prediction error.}
      \label{fig:drivers}
\end{figure*}

\begin{figure}[!h]
\centering
 \includegraphics[width=0.45\textwidth]{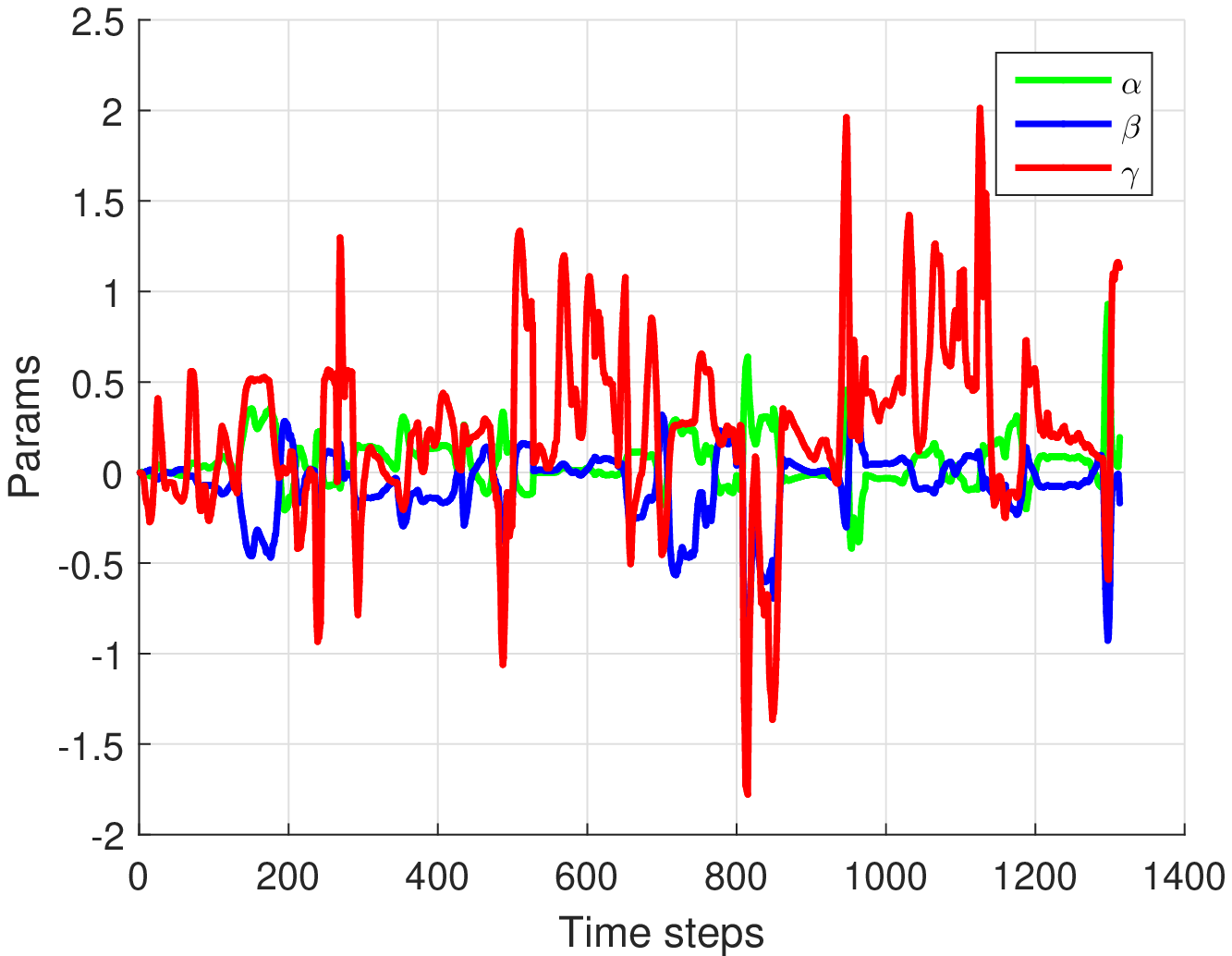}
      \caption{Online estimated parameters.}
      \label{fig:online}
\end{figure}

To validate the proposed model, we use the naturalistic driving data from five randomly selected participants. For each driver, we extract around five car-following episodes where the cruise control is disengaged and there is no relative distance jumps due to lead vehicle lane change or other vehicle cut-ins. We apply the parallel RLS-IQR algorithm for online acceleration prediction for delays varying from 2 steps to 10 steps. The relative distance, vehicle speed, and relative speed are scaled by 1/40, 1/30, and 1/4, respectively to make the three inputs at similar level. A sample trajectory and the prediction with best estimated delays from each driver is shown in Figure~\ref{fig:drivers}. It can be seen that the online prediction offers promising prediction performance. The prediction error statistics for the five drivers are listed in Table~\ref{tab:2}. The online identified parameters for driver 1 are shown in Figure~\ref{fig:online}.

\begin{table}[!h]
\caption{Prediction performance summary}
\label{tab:2}
\centering
\begin{tabular}{c|c|c|c}
\hline
 Driver ID &\# of trips &avg. RMSE &worst RMSE \\
 \hline
1 &9 &0.28 &0.33 \\
\hline
2 &3 &0.32 &0.42\\
\hline
3 &5 &0.26 &0.35 \\
\hline
4 &3 &0.46 &0.52 \\
\hline
5 &2 &0.42 &0.48\\
\hline
\hline
\end{tabular}
\end{table}

\section{Conclusions}\label{sec:6}
In this paper, we developed a novel mechanical-system inspired microscopic traffic model using a mass-spring-damper-clutch system. This model naturally captures general CF behaviors and offers physical interpretations of the CF dynamics. It also considers the impact of the following vehicle on the lead vehicle, which is neglected in existing microscopic CF models and causes issues when chaining vehicles for macroscopic traffic modeling. We develop a parallel recursive least square with inverse QR decomposition algorithm for online parameter identification. The parameter identification has been validated in both simulations and on naturalistic driving data with promising performance.

Future work will include the consideration of nonlinear spring/damper to further improve the model. We will also chain the vehicles together and study the wave propagation on the vehicle platoons.

\begin{IEEEbiography}[{\includegraphics[width=25mm,height=32mm,clip,keepaspectratio]{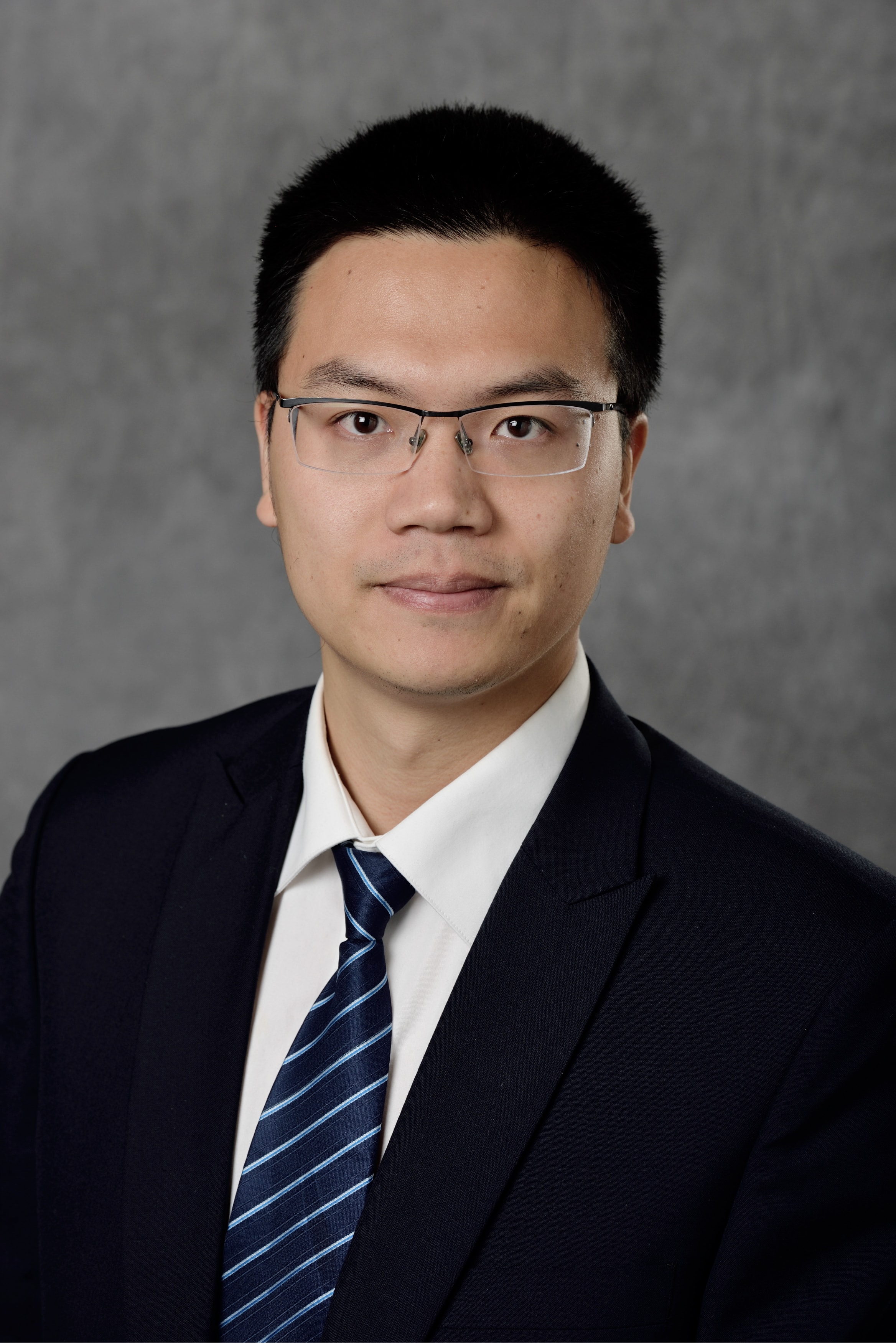}}]{Zhaojian}
received the B.S. degree in civil aviation from the Nanjing University of Aeronautics and Astronautics, Nanjing, China, in 2010, and the M.S. and Ph.D degree from the Department of Aerospace Engineering, University of Michigan, Ann Arbor, MI, USA, in 2014 and 2016, respectively.
From 2010 to 2012, he was an Air Traffic Controller with the Shanghai Area Control Center, Shanghai, China. From 2014 and 2015, he was an Intern with Ford Motor Company, Dearborn, MI, USA. He is currently an Assistant Professor in the Department of Mechanical Engineering, Michigan State University. His current research interests include optimal control, autonomous vehicles, and intelligent transportation systems.
\end{IEEEbiography}

\begin{IEEEbiography}[{\includegraphics[width=25mm,height=32mm,clip,keepaspectratio]{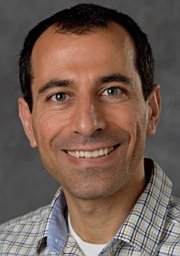}}]{
Firas Khasawneh} received his B.S. degree in mechanical engineering from Jordan University for Science and Technology in Irbid, Jordan, in 2004. In 2007 he received his M.S. in mechanical and aerospace engineering from the University of Missouri in Columbia, while in 2010 he obtained his PhD in Mechanical engineering from Duke University in Durham, North Carolina.
From 2011 until 2013, he was a visiting assistant professor at Duke’s Pratt school of engineering. In 2013, Khasawneh joined SUNY Polytechnic Institute in Utica, New York as an assistant professor of mechanical engineering, and he remained there until 2017 when he joined the mechanical engineering department at Michigan State University as an assistant professor. His current research interests include complex and delayed dynamical systems, time series analysis, machine learning, nonlinear dynamics, and machining dynamics.
\end{IEEEbiography}

\begin{IEEEbiography}[{\includegraphics[width=25mm,height=32mm,clip,keepaspectratio]{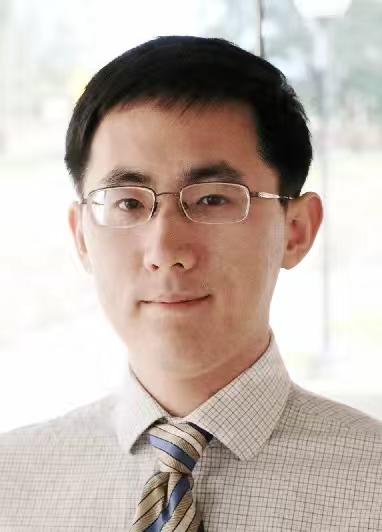}}]{Xiang Yin} was born in Anhui, China, in 1991.
He received the B.Eng degree from Zhejiang University in 2012, the M.S. degree from the University of Michigan, Ann Arbor, in 2013,
and the Ph.D degree from the University of Michigan, Ann Arbor, in 2017, all in electrical engineering.

Since 2017, he has been with the Department of Automation, Shanghai Jiao Tong University, where he is an Associate Professor.
His research interests include formal methods,  control of discrete-event systems, model-based fault diagnosis, security and their applications to cyber and cyber-physical systems.
Dr.\ Yin received the Outstanding Reviewer Awards from \textsc{Automatica}, the \textsc{IEEE Transactions on Automatic Control}  and the \textsc{Journal of Discrete Event Dynamic Systems}.
Dr.\ Yin also received the IEEE Conference on Decision and Control (CDC) Best Student Paper Award Finalist in 2016.
He is the co-chair of the IEEE CSS Technical Committee on Discrete Event Systems.
\end{IEEEbiography}

\begin{IEEEbiography}[{\includegraphics[width=25mm,height=32mm,clip,keepaspectratio]{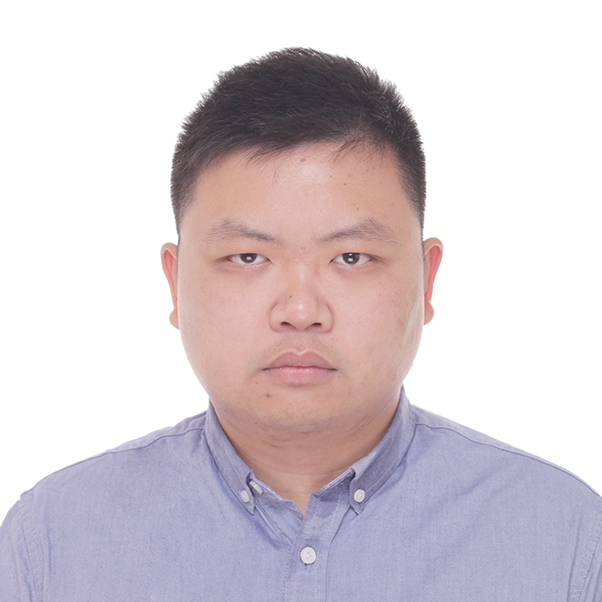}}]{
Aoxue Li} received the B.S. and M.S. degrees in vehicle engineering from Jiangsu University, Zhenjiang, China, in 2013 and 2016, respectively. He is currently pursuing the Ph.D. degree in vehicle engineering at Jiangsu University, Zhenjiang, China. From August 2018 to the present, he is a visiting scholar with the Department of Mechanical Engineering of Michigan State University. His research interests include the autonomous vehicle, intelligent transportation system, and ADAS technologies.
\end{IEEEbiography}

\begin{IEEEbiography}[{\includegraphics[width=25mm,height=32mm,clip,keepaspectratio]{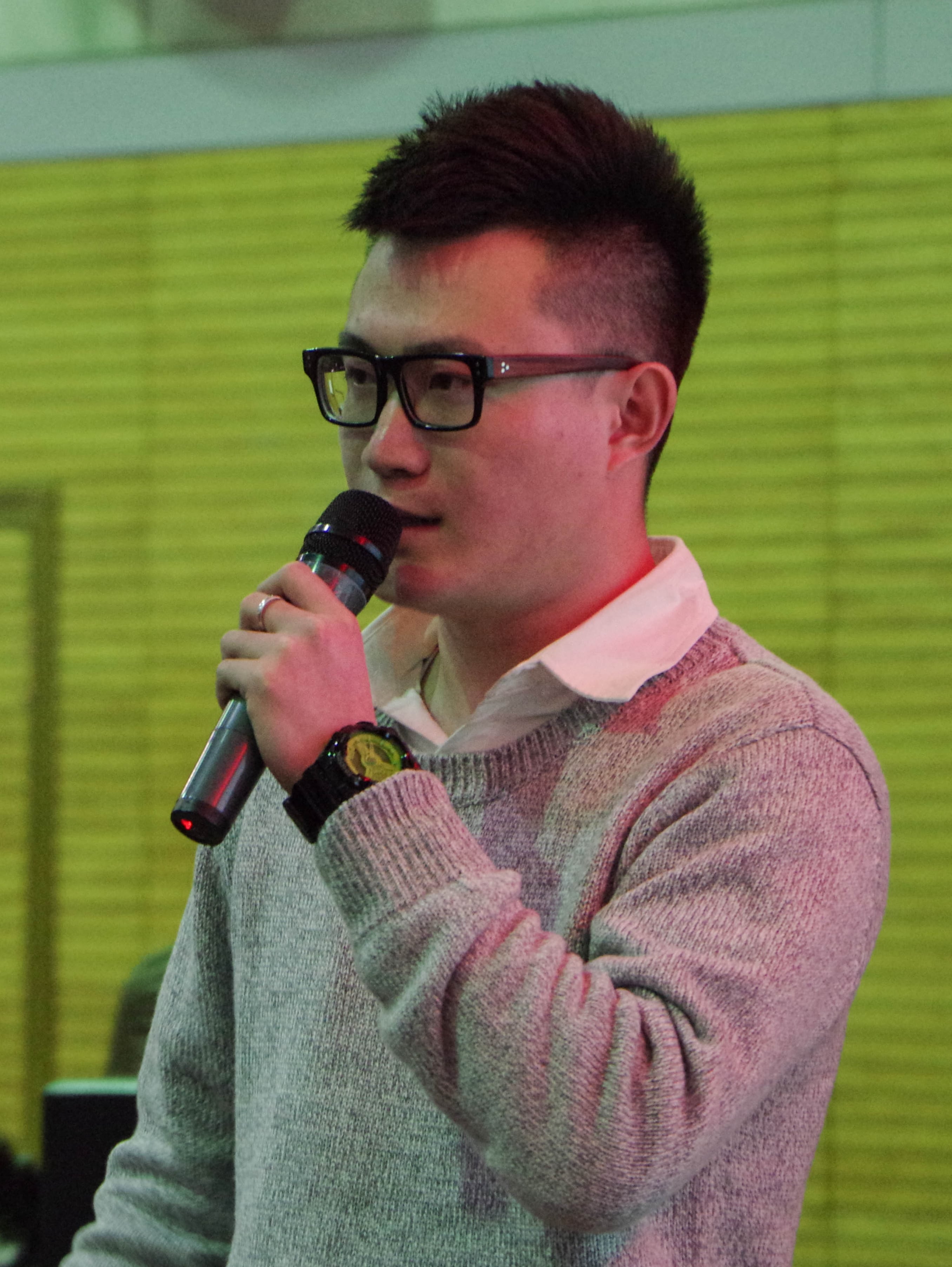}}]{
Ziyou Song} received his B.E. degree (with honor) and the Ph.D. degree (with highest honor) in Automotive Engineering from Tsinghua University, Beijing, China, in 2011 and 2016, respectively. He is currently a post doctor with the University of Michigan, Ann Arbor. His research interests include battery parameter estimation, hybrid energy storage system, electric and hybrid electric vehicles.
\end{IEEEbiography}

\bibliographystyle{ieeetr}
\bibliography{ZL,car_following}

\begin{thebibliography}{10}

\bibitem{inrix}
INRIX, ``Inrix global traffic scorecard,'' 2018.

\bibitem{driving_stress}
D.~A. HENNESSY and D.~L. WIESENTHAL, ``The relationship between traffic
  congestion, driver stress and direct versus indirect coping behaviours,''
  {\em Ergonomics}, vol.~40, no.~3, pp.~348--361, 1997.

\bibitem{pollution}
K.~Zhang and S.~Batterman, ``Air pollution and health risks due to vehicle
  traffic.,'' {\em The Science of the total environment}, vol.~450-451,
  pp.~307--16, 2013.

\bibitem{variable_speed1}
E.~R. Müller, R.~C. Carlson, W.~Kraus, and M.~Papageorgiou, ``Microsimulation
  analysis of practical aspects of traffic control with variable speed
  limits,'' {\em IEEE Transactions on Intelligent Transportation Systems},
  vol.~16, pp.~512--523, Feb 2015.

\bibitem{variable_speed2}
A.~Hegyi, B.~D. Schutter, and J.~Hellendoorn, ``Optimal coordination of
  variable speed limits to suppress shock waves,'' {\em IEEE Transactions on
  Intelligent Transportation Systems}, vol.~6, pp.~102--112, March 2005.

\bibitem{traffic_light1}
M.~B. Younes and A.~Boukerche, ``Intelligent traffic light controlling
  algorithms using vehicular networks,'' {\em IEEE Transactions on Vehicular
  Technology}, vol.~65, pp.~5887--5899, Aug 2016.

\bibitem{traffic_light2}
J.~L. Fleck, C.~G. Cassandras, and Y.~Geng, ``Adaptive quasi-dynamic traffic
  light control,'' {\em IEEE Transactions on Control Systems Technology},
  vol.~24, pp.~830--842, May 2016.

\bibitem{metering1}
Y.~Wang, E.~B. Kosmatopoulos, M.~Papageorgiou, and I.~Papamichail, ``Local ramp
  metering in the presence of a distant downstream bottleneck: Theoretical
  analysis and simulation study,'' {\em IEEE Transactions on Intelligent
  Transportation Systems}, vol.~15, pp.~2024--2039, Oct 2014.

\bibitem{metering2}
I.~Papamichail and M.~Papageorgiou, ``Traffic-responsive linked ramp-metering
  control,'' {\em IEEE Transactions on Intelligent Transportation Systems},
  vol.~9, pp.~111--121, March 2008.

\bibitem{Macro}
C.~F. Daganzo, ``The cell transmission model, part ii: Network traffic,'' {\em
  Transportation Research Part B: Methodological}, vol.~29, no.~2, pp.~79 --
  93, 1995.

\bibitem{Kinematic1}
``Shock waves on the highway,'' {\em Oper. Res.}, vol.~4, pp.~42--51, Feb.
  1956.

\bibitem{Kinematic2}
C.~F. Daganzo, ``A finite difference approximation of the kinematic wave model
  of traffic flow,'' {\em Transportation Research Part B: Methodological},
  vol.~29, no.~4, pp.~261 -- 276, 1995.

\bibitem{Kinematic3}
D.~Helbing, A.~Hennecke, V.~Shvetsov, and M.~Treiber, ``Master: macroscopic
  traffic simulation based on a gas-kinetic, non-local traffic model,'' {\em
  Transportation Research Part B: Methodological}, vol.~35, no.~2, pp.~183 --
  211, 2001.

\bibitem{Dynamic1}
B.~S. Kerner and P.~Konh\"auser, ``Cluster effect in initially homogeneous
  traffic flow,'' {\em Phys. Rev. E}, vol.~48, pp.~R2335--R2338, Oct 1993.

\bibitem{Dynamic2}
H.-N. Nguyen, B.~Fishbain, E.~Bitar, D.~Mahalel, and P.-O. Gutman, ``Dynamic
  model for estimating the macroscopic fundamental diagram,'' {\em
  IFAC-PapersOnLine}, vol.~49, no.~3, pp.~297 -- 302, 2016.
\newblock 14th IFAC Symposium on Control in Transportation SystemsCTS 2016.

\bibitem{Hydro1}
R.~Kaur and S.~Sharma, ``Analyses of a heterogeneous lattice hydrodynamic model
  with low and high-sensitivity vehicles,'' {\em Physics Letters A}, vol.~382,
  no.~22, pp.~1449 -- 1455, 2018.

\bibitem{Hydro2}
H.~Ge, R.~Cheng, and L.~Lei, ``The theoretical analysis of the lattice
  hydrodynamic models for traffic flow theory,'' {\em Physica A: Statistical
  Mechanics and its Applications}, vol.~389, no.~14, pp.~2825 -- 2834, 2010.

\bibitem{CA1}
E.~Kometani and T.~Sasaki, ``Dynamic behaviour of traffic with a nonlinear
  spacing-speed relationship,'' in {\em Proceedings of the Symposium on Theory
  of Traffic Flow}, pp.~105--119, 1959.

\bibitem{CA2}
P.~Gipps, ``A behavioural car-following model for computer simulation,'' {\em
  Transportation Research Part B: Methodological}, vol.~15, no.~2, pp.~105 --
  111, 1981.

\bibitem{1950}
L.~A. Pipes, ``An operational analysis of traffic dynamics,'' {\em Journal of
  Applied Physics}, vol.~24, no.~3, pp.~274--281, 1953.

\bibitem{GHR}
D.~Gazis, R.~Herman, and R.~Rothery, ``Nonlinear follow-the-leader models of
  traffic flow,''

\bibitem{review}
M.~Brackstone and M.~McDonald, ``Car-following: a historical review,'' {\em
  Transportation Research Part F: Traffic Psychology and Behaviour}, vol.~2,
  no.~4, pp.~181 -- 196, 1999.

\bibitem{GHR1}
R.~Herman and R.~B. Potts, ``Single lane traffic theory and experiment,'' in
  {\em Proceedings of the Symposium on Theory of Traffic Flow}, pp.~147--157,
  Elsevier, 1959.

\bibitem{GHR2}
A.~May and H.~Keller, ``Non integer car following models,'' {\em Highway
  Research Record}, vol.~199, pp.~19--32, 1967.

\bibitem{GHR3}
H.~Ozaki, ``Reaction and anticipation in the car following behavior,'' in {\em
  Proceedings of the 13th International Symposium on Traffic and Transportation
  Theory}, pp.~349--366, 1993.

\bibitem{Helly}
W.~Helly, ``Simulation of bottlenecks in single lane traffic flow,'' in {\em
  Proceedings of Symposium on Traffic and Transportation Theory}, pp.~207--238,
  1959.

\bibitem{Helly1}
M.~Aron, ``Car following in an urban network: simulation and experiments,'' in
  {\em Proceedings of Seminar D, 16th {PTRC} Meeting}, pp.~27--39, 1988.

\bibitem{Helly2}
J.~Xing, ``A parameter identification of a car following model,'' in {\em
  Proceedings of the Second World Congress on {ATT}}, pp.~1739--1745, 1995.

\bibitem{AP1}
R.~Michaels, ``Perceptual factors in car following,'' in {\em Proceedings of
  the Second International Symposium on the Theory of Road Traffic Flow},
  pp.~44--59, 1963.

\bibitem{AP2}
L.~Evans and R.~Rothery, ``Experimental measurement of perceptual thresholds in
  car following,'' {\em Highway Research Record}, vol.~64, pp.~13 -- 29, 1973.

\bibitem{FL}
C.~Kikuchi and P.~Chakroborty, ``Car following model based on a fuzzy inference
  system,'' {\em Transportation Research Record}, vol.~1365, pp.~82--91, 1992.

\bibitem{spring}
Y.~Li, W.~Chen, S.~Peeta, X.~He, T.~Zheng, and H.~Feng, ``An extended
  microscopic traffic flow model based on the spring-mass system theory,''
  {\em Modern Physics Letters B}, vol.~31, no.~09, p.~1750090, 2017.

\bibitem{Khasawneh2013}
F.~A. Khasawneh and B.~P. Mann, ``A spectral element approach for the stability
  analysis of time-periodic delay equations with multiple delays,'' {\em
  Communications in Nonlinear Science and Numerical Simulation}, vol.~18,
  pp.~2129--2141, aug 2013.

\bibitem{jin}
J.~I. Ge and G.~Orosz, ``Connected cruise control among human-driven vehicles:
  Experiment-based parameter estimation and optimal control design,'' {\em
  Transportation Research Part C: Emerging Technologies}, vol.~95, pp.~445 --
  459, 2018.

\bibitem{IQR}
S.~T. Alexander and A.~L. Ghimikar, ``A method for recursive least squares
  filtering based upon an inverse qr decomposition,'' {\em IEEE Transactions on
  Signal Processing}, vol.~41, pp.~20--, January 1993.

\bibitem{Insperger2004}
T.~Insperger and G.~St{\'{e}}p{\'{a}}n, ``Updated semi-discretization method
  for periodic delay-differential equations with discrete delay,'' {\em
  International Journal for Numerical Methods in Engineering}, vol.~61,
  pp.~117--141, aug 2004.

\bibitem{Butcher2004}
E.~A. Butcher, H.~Ma, E.~Bueler, V.~Averina, and Z.~Szabo, ``Stability of
  linear time-periodic delay-differential equations via chebyshev
  polynomials,'' {\em International Journal for Numerical Methods in
  Engineering}, vol.~59, pp.~895--922, jan 2004.

\bibitem{Tweten2012}
D.~J. Tweten, G.~M. Lipp, F.~A. Khasawneh, and B.~P. Mann, ``On the comparison
  of semi-analytical methods for the stability analysis of delay differential
  equations,'' {\em Journal of Sound and Vibration}, vol.~331, pp.~4057--4071,
  aug 2012.

\bibitem{CCC}
J.~I. Ge and G.~Orosz, ``Connected cruise control among human-driven vehicles:
  Experiment-based parameter estimation and optimal control design,'' {\em
  Transportation Research Part C: Emerging Technologies}, vol.~95, pp.~445 --
  459, 2018.

\bibitem{CCC1}
G.~Orosz, ``Connected cruise control: modelling, delay effects, and nonlinear
  behaviour,'' {\em Vehicle System Dynamics}, vol.~54, no.~8, pp.~1147--1176,
  2016.

\bibitem{CCC2}
G.~Orosz, R.~E. Wilson, and G.~St{\'e}p{\'a}n, ``Traffic jams: dynamics and
  control,'' {\em Philosophical Transactions of the Royal Society of London A:
  Mathematical, Physical and Engineering Sciences}, vol.~368, no.~1928,
  pp.~4455--4479, 2010.

\bibitem{IVBSS}
J.~Sayer, D.~LeBlanc, S.~Bogard, D.~Funkhouser, S.~Bao, M.~Buonarosa, and
  A.~Blankespoor, ``Integrated vehicle-based safety systems field operational
  test final program report,'' tech. rep., 2011.

\bibitem{Eye}
G.~P. Stein, E.~Rushinek, G.~Hayun, and A.~Shashua, ``A computer vision system
  on a chip: a case study from the automotive domain,'' in {\em 2005 IEEE
  Computer Society Conference on Computer Vision and Pattern Recognition
  (CVPR'05) - Workshops}, pp.~130--130, Sept 2005.

\end{thebibliography}
\end{document}